\documentclass[12pt,preprint,letterpaper]{aastex}
\usepackage{url}
\usepackage{hyperref}
\usepackage{amsmath}
\usepackage{tabu}
\usepackage{booktabs}
\usepackage{todonotes}
\usepackage{comment}
 \usepackage{color}           
 \definecolor{DarkGreen}{rgb}{0.0,0.45,0.0}  

\newcommand{\fig}[1]{Figure~\ref{#1}}
\newcommand{\tbl}[1]{Table~\ref{#1}}

\newcommand{\uvec}[1]{ \widehat{\mathbf{#1}} }

\newcommand{\parb}{ \frac{\partial\uvec{B}}{\partial\mathbf{r}} }

\begin{document}

\shorttitle{Magnetic Twist} %

\shortauthors{Liu et al.}

\title{Structure, Stability, and Evolution of Magnetic Flux Ropes from the Perspective of Magnetic Twist} 

\author{Rui Liu\altaffilmark{1,2}, Bernhard Kliem\altaffilmark{3,4}, Viacheslav S. Titov\altaffilmark{5}, Jun Chen\altaffilmark{1,6}, Yuming Wang\altaffilmark{1,2}, Haimin Wang\altaffilmark{7,8}, Chang Liu\altaffilmark{7,8}, Yan Xu\altaffilmark{7,8}, Thomas Wiegelmann\altaffilmark{9}}

\altaffiltext{1}{CAS Key Laboratory of Geospace Environment, Department of Geophysics and Planetary Sciences, University of Science and
Technology of China, Hefei 230026, China}

\altaffiltext{2}{Collaborative Innovation Center of Astronautical Science and Technology, China}

\altaffiltext{3}{Yunnan Observatories, CAS, Kunming 650216, Hefei 230026, China}

\altaffiltext{4}{Institute of Physics and Astronomy, University of Potsdam, 14476 Potsdam,
Germany}

\altaffiltext{5}{Predictive Science Inc., 9990 Mesa Rim Road, Suite 170, San Diego, CA 92121, USA}
 
\altaffiltext{6}{Mengcheng National Geophysical Observatory, School of 
 	Earth and Space Sciences, University of Science and Technology of 
 	China, Hefei 230026, China}
          
\altaffiltext{7}{Space Weather Research Laboratory, New Jersey Institute of Technology, University Heights, Newark, NJ 07102-1982, USA}

\altaffiltext{8}{Big Bear Solar Observatory, New Jersey Institute of Technology, 40386 North Shore Lane, Big Bear City, CA 92314-9672, USA}

\altaffiltext{9}{Max-Planck-Institut f\"{u}r Sonnensystemforschung, Justus-von-Liebig-Weg 3, 37077 G\"{o}ttingen, Germany}

\email{rliu@ustc.edu.cn}

\begin{abstract}
We investigate the evolution of NOAA Active Region 11817 during 2013 August 10--12, when it developed a complex field configuration and produced four confined, followed by two eruptive, flares. These C-and-above flares are all associated with a magnetic flux rope (MFR) located along the major polarity inversion line, where shearing and converging photospheric flows are present. Aided by the nonlinear force-free field modeling, we identify the MFR through mapping magnetic connectivities and computing the twist number $\mathcal{T}_w$ for each individual field line. The MFR is moderately twisted ($|\mathcal{T}_w| < 2$) and has a well-defined boundary of high squashing factor $Q$. We found that the field line with the extremum $|\mathcal{T}_w|$ is a reliable proxy of the rope axis, and that the MFR's peak $|\mathcal{T}_w|$ temporarily increases within half an hour before each flare while it decreases after the flare peak for both confined and eruptive flares. This pre-flare increase in $|\mathcal{T}_w|$ has little effect on the active region's free magnetic energy or any other parameters derived for the whole region, due to its moderate amount and the MFR's relatively small volume, while its decrease after flares is clearly associated with the stepwise decrease in the whole region's free magnetic energy due to the flare. We suggest that $\mathcal{T}_w$ may serve as a useful parameter in forewarning the onset of eruption, and therefore, the consequent space weather effects. The helical kink instability is identified as the prime candidate onset mechanism for the considered flares.
\end{abstract}


\section{Introduction}
It is generally accepted that solar storms including flares, filament eruptions, and coronal mass ejections (CMEs) are manifestations of an explosive release of free magnetic energy stored in corona, involving large-scale disruption and restructuring of the coronal magnetic field. Major contributors to the accumulation of coronal magnetic energy are 1) emergence of magnetic flux through the photosphere, and 2) small- and large-scale convective motions in the photospheric layers that shuffle around the footpoints of the coronal field lines \citep{wiegelmann12}. These two processes also transport magnetic helicity \citep{berger84} into the solar atmosphere. Typically a higher content of free magnetic energy and relative helicity is found in active regions that are more flare productive or prior to more energetic flares \citep{tw08, jing09, sun12, maeshiro05, labonte07, park08, park10, Tziotziou12}. It is an appealing conjecture that there may exist a critical amount of helicity, above which an eruptive process is favored \citep{Zhang&Flyer08, park08, park10, Tziotziou12, Tziotziou13}. This conjecture, however, is challenged by numeric experiments \citep{phillips05,Zuccarello09}. Further, the accumulation of sufficient energy may not guarantee the occurrence of eruptions \citep[e.g.,][]{Gilchrist12}. Overall, the pre-eruptive state of the coronal magnetic field and the conditions that signal impending eruptions remain elusive. 

To understand the initiation of solar eruptions, both magnetic reconnection \citep[e.g.,][]{moore01, adk99} and ideal MHD instabilities \citep[e.g.,][]{hood&priest79, vanTend&Kuperus1978, forbes&priest95, kliem&torok06} have been under intense investigation. In particular, the helical kink instability has received a lot of attention \citep[e.g.,][]{fan05, torok04, torok&kliem05, kliem10}, mainly motivated by the dramatic development of writhe in eruptive structures \citep[e.g.,][]{ji03, romano03, rust&labonte05, williams05, alexander06, liu07, liu&alexander09, cho09, karlicky&kliem10, liu12, kumar12, yang12, kumar&cho14}, most of which are filaments. These events exhibit both a winding of the filament threads about the axis, arguing for the existence of considerable twist, and an overall helical shape, indicating a writhed axis. This combination therefore strongly indicates the kink instability of a magnetic flux rope (MFR), whereby magnetic twist (winding of magnetic field lines around the rope axis) is abruptly converted to magnetic writhe (winding of the axis itself). The instability is triggered when the twist exceeds a threshold, whose precise value depends on the details of the configuration. With photospheric line tying included, the minimum threshold is a winding of the field lines about the rope axis by 1.25 turns \citep[e.g.,][]{hood&priest81, einaudi&vanhoven83, baty&heyvaerts96, torok&kliem03}.

However, it has been under debate whether the kink instability plays a significant role in solar eruptions. A previous question on the sufficiency of twist in active regions \citep{leamon03} was apparently settled by examining the magnetic fields in localized active-region MFRs \citep{leka05}, but there are more issues for consideration. First of all, eruptive structures with a clear writhing feature are relatively rare, which raises a question as to how often the kink instability triggers eruptions. Second, twisted, helical patterns are often exposed only during eruptions \citep[e.g.,][]{vrsnak91, vrsnak93, romano03, gary&moore04, srivastava10, kumar12}. It is hence difficult to determine whether the twist is accumulated prior to the eruption or built up in the course of the eruption through reconnection in the vertical current sheet under the rope \citep{lin03}. \citet{qiu07} found that the magnetic reconnection flux in the low corona in the wake of CMEs is comparable to that in the resultant interplanetary magnetic clouds, suggesting the formation of the helical structure of MFRs by reconnection. This raises a question as to whether the observed kink is actually a byproduct of the eruption. Third, the shear component of the ambient field may cause a similar writhing of the current-carrying MFR as the kink mode \citep{isenberg&forbes07}. This may be excluded as an important process only in a few cases with Dopplergrams available, where both the writhing motion in the early phase of the eruption and the un-writhing relaxation later on are detected \citep{alexander06,liu07}, and in the few cases of extremely strong writhing resulting in apex rotations significantly higher than 90 degrees \citep{Kliem2012}. And finally, hard X-ray and microwave emission at the projected crossing point of kinked filaments \citep{alexander06, liu&alexander09, cho09, karlicky&kliem10} suggests that the their legs approach each other and interact near the crossing point. This is possible when the MFR is highly twisted \citep{kliem10}. Occasionally, high twist values are estimated for eruptive structures \citep[e.g.,][]{vrsnak91, vrsnak93, romano03, gary&moore04, srivastava10, kumar12}, but such estimates suffer inevitably from projection effects and large uncertainties. In contrast, nonlinear force-free field (NLFFF) modeling often yields weakly twisted MFRs associated with pre-eruptive or quiescent filaments \citep{ra04,canou09, bobra08, jing10, guo10b, guo10a, YGuo&al2013, su11}, in most cases with a twist below 1.5 turns. Furthermore, numerical simulations have demonstrated that it is difficult for an MFR to rise coherently into the corona owing to the dense plasma trapped at the concave-upward portions of the twisted field lines \citep[][and references therein]{magara06}. Hence, one may hypothesize that the majority of pre-eruptive MFRs in corona are only moderately or weakly twisted. 

Here we present a study of the evolution of MFRs in relation to flares/CMEs in NOAA Active Region (AR) 11817, which sheds light on the role of the kink instability in the flare/CME production. In this paper, an MFR is defined as a collection of magnetic field lines spiraling around the same axis by more than one full turn. By this rigorous definition, MFRs are identified without ambiguity through the NLFFF modeling. It has been demonstrated that the NLFFF extrapolation is capable of reconstructing MFRs and other topological features of active region magnetic fields, with high fidelity \citep[e.g.,][]{valori05, wiegelmann06aa, valori10}. In the sections that follow, our methods of data reduction are described in Section \ref{sec:data}, the observations of the MFRs are presented in Section \ref{sec:observation}, and a discussion and concluding remarks are given in Section \ref{sec:conclusion}. 

\section{Instruments and Data Reduction} \label{sec:data}
\subsection{Instruments}
For this study we used vector magnetograms obtained by the Helioseismic and Magnetic Imager \citep[HMI;][]{hoeksema14} onboard the Solar Dynamics Observatory \citep[SDO;][]{pesnell12}. The data for HMI Active Region Patches (HARPs) are disambiguated and deprojected to the heliographic coordinates with a Lambert (cylindrical equal area) projection method, resulting in a pixel scale of $0.03^\circ$ or 0.36 Mm \citep{Bobra2014}. \citet{liuy14} have demonstrated that different projection methods make little impact on the deprojected map of a normal AR and on the calculation of the helicity flux. 

The flares produced by AR 11817 were observed with the Atmospheric Imaging Assembly \citep[AIA;][]{lemen12} onboard SDO. The six EUV passbands of AIA, i.e., 131~{\AA} (peak response temperature $\log T=7.0$), 94~{\AA} ($\log T=6.8$), 335~{\AA} ($\log T=6.4$), 211~{\AA} ($\log T=6.3$), 193~{\AA} ($\log T=6.1$) and 171~{\AA} ($\log T=5.8$), can be used to calculate the differential emission measure (DEM) in the logarithmic temperature range $[5.5,7.5]$. We utilized the regularized inversion code developed by \citet{hk12} to recover DEMs, using the most recent available AIA temperature response functions. The AIA level-1 data are further processed by applying the routines \texttt{AIA\_DECONVOLVE\_RICHARDSONLUCY} and \texttt{AIA\_PREP} from the SolarSoft package before being fed into the DEM code. 

One of the flares produced by AR 11817 is captured by the 1.6-m New Solar Telescope (NST) at Big Bear Solar Observatory (BBSO) with unprecedented high spatiotemporal resolution. The telescope adopts a state-of-art high-order adaptive optics system of 308 elements to correct atmospheric disturbances. Diffraction-limited imaging is achieved with the aid of speckle image reconstruction. The spatial resolution of the H$\alpha$ images used in this study is $0.085''$ (60 km) at a cadence of 15 s \citep[see][for more detail]{wang15}.

\subsection{Helicity \& Energy Input}
To monitor the evolution and complexity of the AR, we calculated the relative helicity flux across its photospheric boundary $S$ with the following formula \citep{berger84,liuy12,liuy14}:
\begin{equation}
\left.\frac{dH}{dt}\right|_S=2\int_S(\mathbf{A}_p\cdot\mathbf{B}_t)V_{\perp n}\,dS-2\int_S(\mathbf{A}_p\cdot\mathbf{V}_{\perp t})B_n\,dS,
\end{equation}
where $\mathbf{A}_p$ is the vector potential of the reference potential field $\mathbf{B}_p$ that has the same vertical component on the photospheric boundary; $t$ and $n$ refer to the tangential and normal directions, respectively. $\mathbf{V}_\perp$ is the photospheric velocity that is perpendicular to magnetic field lines. To obtain $\mathbf{V}_\perp$, we applied the Differential Affine Velocity Estimator for Vector Magnetograms \citep[DAVE4VM;][]{schuck08} to the time-series of deprojected, registered vector magnetograms. We then subtracted the field-aligned plasma flow $(\mathbf{V}\cdot\mathbf{B})\mathbf{B}/B^2$ from the velocity derived by DAVE4VM to yield $\mathbf{V}_\perp$ \citep{liuy12}. The window size used in DAVE4VM is chosen to be 19 pixels, following \citet{liuy14}. The first term in the above equation represents helicity injection due to the emergence of twisted flux tubes into the corona; the second is due to photospheric motions that shear and braid field lines. Similarly, both flux emergence ($V_{\perp n}$) and tangential motions ($\mathbf{V}_{\perp t}$) contribute to the Poynting flux across the photospheric boundary $S$, which is given by \citep{kusano02}
\begin{equation}
\left.\frac{dP}{dt}\right|_S=\frac{1}{4\pi}\int_S B_t^2V_{\perp n}\,dS-\frac{1}{4\pi}\int_S(\mathbf{B}_t\cdot\mathbf{V}_{\perp t})B_n\,dS.
\end{equation} Readers are referred to Appendix~\ref{append:error} for an error analysis of helicity and Poynting flux.

\subsection{Map of Magnetic Connectivities}
To understand the magnetic connectivities within the active region and their evolution, we used the code package developed by T.~Wiegelmann, which utilizes the ``weighted optimization'' method \citep{wiegelmann04,wiegelmann12} to build an NLFFF model that approximates the coronal field (see Appendix~\ref{append:quality} for the quality of NLFFF). To best suit the force-free condition, the vector magnetograms are ``pre-processed'' \citep{wiegelmann06} before being taken as the photospheric boundary. Our calculation is performed within a box of $512\times 256\times 256$ uniformly spaced grid points, whose photospheric FOV is shown in Figure~\ref{bfield}. Further, the free magnetic energy accumulated in the active region can be derived by subtracting the magnetic energy of the corresponding potential field from that of the NLFFF (see Appendix~\ref{append:error} for an error analysis of magnetic free energy). The potential field is calculated by the Green function method. We calculated the free magnetic energy regularly on an hourly cadence, but at the highest cadence available (12 min) around flares. It is clear that the NLFFF is not a good model of the coronal active-region field during the impulsive phase of the flares, as plasma is accelerated, i.e., forces are significant, primarily in this phase \citep{JZhang&al2001}. However, the flare-related changes of the coronal field can be inferred from a comparison of the NLFFF before and after the impulsive flare phase.

After obtaining the NLFFF, we refined the photospheric computational grid by 16 times and traced magnetic field lines pointwise with a fourth-order Runge-Kutta method to ensure high precision. The mapped footpoints of field lines were used to calculate the squashing factor $Q$ of elemental magnetic flux tubes \citep{titov02,titov07}.
Basically, for a mapping defined by the two footpoints of a field line $\Pi_{12}:  \mathbf{r}_1(x_1,\ y_1)\mapsto \mathbf{r}_2(x_2,\ y_2)$, the Jacobian matrix associated with the mapping is 
\begin{equation} \label{eq:d12} 
D_{12}=\left[\frac{\partial \mathbf{r}_2}{\partial \mathbf{r}_1}\right]= 
\begin{pmatrix}
\partial x_2/\partial x_1 &\partial x_2/\partial y_1 \\
\partial y_2/\partial x_1 & \partial y_2/\partial y_1
\end{pmatrix} \equiv 
\begin{pmatrix}
a & b \\
c & d
\end{pmatrix},
\end{equation}
and then the squashing factor associated with the field line is defined as \citep{titov02}
\begin{equation} \label{eq:q}
Q\equiv \frac{a^2+b^2+c^2+d^2}{|B_{n,1}(x_1,y_1)/B_{n,2}(x_2,y_2)|},
\end{equation}
where $B_{n,1}(x_1,y_1)$ and $B_{n,2}(x_2,y_2)$ are the components normal to the plane of the footpoints, in our case, the photosphere, and their ratio is equivalent to the determinant of $D_{12}$. Quasi-separatrix layers (QSLs) as defined by high-$Q$ values are often complex three-dimensional structures, hence their visualization can be facilitated by $Q$-maps cutting across the QSL of interest. \cite{Pariat&Demoulin12} discussed three relevant methods and found that the method (their Method 3) utilizing the field-line mappings between the cutting plane and the footpoint planes gives optimal results, i.e., the chain rule of the Jacobian is used to calculate $D_{12}$, from a set of field lines threading the cutting plane in a neighborhood of $\mathbf{r}_c(x_c,y_c)$:
\begin{equation}
D_{12}=\left[\frac{\partial {\bf r}_2}{\partial {\bf r}_1}\right] =
 \left[\frac{\partial {\bf r}_2}{\partial {\bf r}_c}\right] \times \left[\frac{\partial {\bf r}_c}{\partial {\bf r}_1}\right] \label{eq:qcs}
\end{equation}
where $(x_1,y_1)$ and $(x_2,y_2)$ are again the corresponding photospheric footpoints and the Jacobian associated with the mapping $\Pi_{1c}:\mathbf{r}_1(x_1,\ y_1)\mapsto \mathbf{r}_c(x_c,\ y_c)$ is given by its inverse, i.e.,  
\begin{equation}
\left[\frac{\partial {\bf r}_c}{\partial {\bf r}_1}\right] = \frac{1}{|B_{n,c}(x_c,y_c)/B_{n,1}(x_1,y_1)|} 
\begin{pmatrix}
\partial y_1/\partial y_c & -\partial x_1/\partial y_c \\
-\partial y_1/\partial x_c & \partial x_1/\partial x_c
\end{pmatrix}.
\end{equation}
We found, however, that field lines touching the cutting plane, i.e., $B_{n,c}(x_c,y_c)\rightarrow 0$, introduce spurious high-$Q$ structures. These can be effectively eliminated via two different approaches: for a field line making a small angle to the cutting plane, one may calculate its $Q$-value from a set of field lines originating from the neighborhood of its photospheric footpoint, which is equivalent to Method 2 in \citet{Pariat&Demoulin12}. One problem is that the new $Q$-values in replacement of the original spurious high-$Q$ structures may be much smaller than those in the neighborhood, calculated with Eq.~\ref{eq:qcs}, therefore giving rise to artificial abrupt changes. The latter are not intrinsic to the magnetic structure under study but rather are due to numerical errors that are generally different for the two methods. Alternatively, one may switch locally to a new plane that is perpendicular\footnote{For convenience, it is sufficient to use a plane perpendicular to the dominant Cartesian component of the local magnetic field vector.} to this particular field line to calculate its $Q$-value again with Eq.~\ref{eq:qcs}. This new value is often consistent with those in the surroundings.

\subsection{Map of Magnetic Twist}
We further calculated the twist number $\mathcal{T}_w$ to measure how many turns two infinitesimally close field lines wind about each other \citep[][Eq.16]{bp06}:  
\begin{eqnarray}
\mathcal{T}_w&=&\int_L\frac{\mu_0J_\parallel}{4\pi B}\,dl = \int_L\frac{\nabla\times\mathbf{B}\cdot\mathbf{B}}{4\pi B^2}\,dl  \label{eq:Tw} \\
&=&\frac{1}{4\pi}\int_L \alpha\,dl, \quad \text{if } \nabla\times\mathbf{B}=\alpha \mathbf{B}, \label{eq:Tw_alpha}
\end{eqnarray} 
where $\alpha$ is the force-free parameter, and $\nabla\times\mathbf{B}\cdot\mathbf{B}/4\pi B^2$ can be regarded as a local density of twist along the field line. In Appendix~\ref{append:tw} we address the relation among $\mathcal{T}_w$, twist number of an individual magnetic field line, $\mathcal{T}_g$, twist number of any curve about an axis in a general geometry \citep[see Eq.12 in][]{bp06}, and $N$, twist number of a field line about the axis of a cylindrical flux tube. Simply put, $\mathcal{T}_g$ is the generalization of $N$, and $\mathcal{T}_w$ approaches $\mathcal{T}_g$ and $N$ in the vicinity of the axis of a nearly cylindrically symmetric flux tube, but deviates otherwise (see Appendix~\ref{append:tw}). While $\mathcal{T}_g$ is the quantity usually considered in stability analyses, its computation depends upon the correct determination of the axis, which is non-trivial for numerically given fields and demanding for a large number of NLFFFs. Therefore, we employ $\mathcal{T}_w$ in the analysis of this paper.

For a perfect NLFFF, $\alpha$ is a constant for each individual field line. In practice, we linearly interpolate $\mathbf{B}$ and $\nabla\times\mathbf{B}$ on the line position and carry out the integration with a 5-point Newton-Cotes formula. We refrained from linearly interpolating $\alpha$ on the non-grid points because of its inherent nonlinearity. Consequently, a map of twist numbers (`twist map' hereafter) is yielded at the photosphere or in a cutting plane by assigning the twist number of each field line to the position where the line threads the plane.  Alternatively, \citet{Inoue2011} obtained the photospheric twist map by multiplying the photospheric maps of $\alpha$ and field line length. To address the issue that $\alpha$ is often not the same at the conjugate footpoints, they took the mean $\alpha$ at the conjugate footpoints for each individual field line \citep[see also][]{Chintzoglou2015}. A comparison between the different approaches to calculate a twist number is given in Appendix~\ref{append:tw}.

The codes that are developed by R.~Liu and J.~Chen to calculate $Q$ and $\mathcal{T}_w$ are available online at \url{http://staff.ustc.edu.cn/~rliu/qfactor.html}. 
 
\section{Observation and Analysis} \label{sec:observation}
We investigated the evolution of AR 11817 during the first three days after its emergence in the southern hemisphere at the beginning of 2013 August 10. Both NLFFF modeling and EUV observations suggest the existence of an MFR at the major polarity inversion line (PIL), where a series of flares and CMEs took place. We concentrated on the MFR and its coronal response, namely, flares, CMEs, and plasma heating. The goal of this study is to demonstrate that an MFR exists prior to each of the flares, to quantify its parameters, and to understand its eruption mechanism.

\begin{figure}
  \centering
  \includegraphics[width=\hsize]{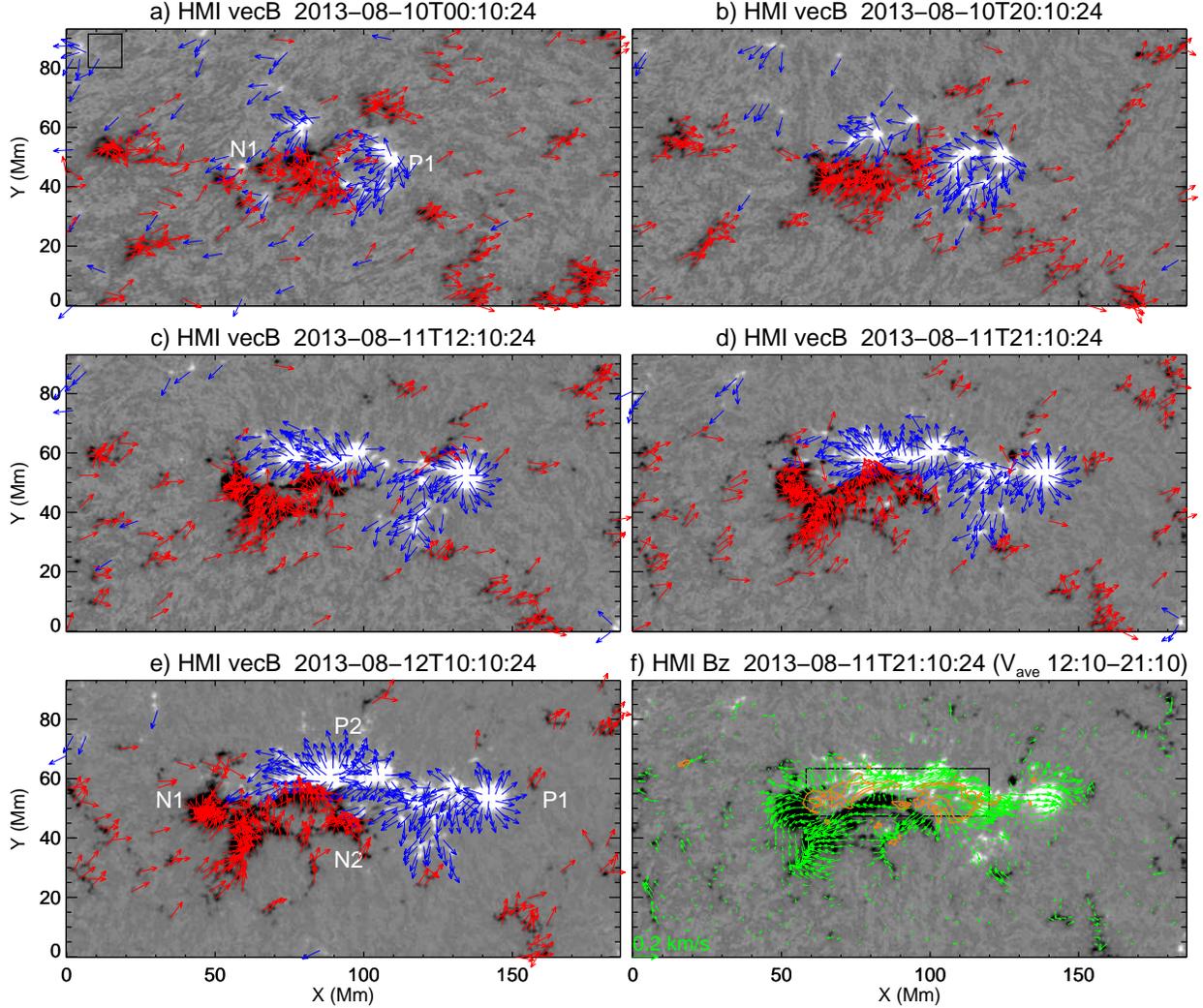}
  \caption{\small Evolution of AR 11817. White and black colors refer to the positive and negative nomal field component, respectively, which are scaled to $\pm800$ G. Red (blue) arrows in Panels (a--e) represent the tangential field component, which originate from negative (positive) normal component whose magnitude exceeds 100 G. In Panel (f), green arrows indicate the tangential velocities ($\mathbf{V}_{\perp t}$), and orange contours refer to normal velocities ($V_{\perp n}$) at 0.05 and 0.08 km s $^{-1}$ (upflows). Only those vectors (tangential field or velocity) at the pixels where the normal field component exceeds 100 G in magnitude are plotted. The velocity field is averaged over 9 hours from 12:10--21:10 UT on 2013 August 11. This time interval corresponds to the period when both the helicity and Poynting fluxes increase rapidly. The rectangular region in Panel (a) is used for error analysis (see Appendix~\ref{append:error}), whereas the rectangle in Panel (f) indicates the FOV of photospheric twist maps and $Q$-maps in this paper. \label{bfield}}
\end{figure}

\subsection{Evolution of AR 11817}
AR 11817 emerged on August 10 as a simple $\beta$ configuration (P1-N1; Figure~\ref{bfield}(a)) with an almost east-west orientation obeying Hale's law. As time progressed, P1-N1 departed from each other and grew in size. Meanwhile, a new bipole P2-N2 emerged on the neutral line of P1-N1, defying Hale's law (Figure~\ref{bfield}(c--e)). There were significant shearing and converging motions, with P2 moving westward and N2 moving northeastward (Figure~\ref{bfield}(f)). As a result the pair exhibited a slow clockwise rotation, which could be explained by the emergence of a magnetic tube with left-handed writhe. The tube might be downward-kinked because of the converging motion, which mainly manifests as N2's moving towards P2. In contrast, no clear sign of writhe is seen for P1-N1. With the emergence of P2-N2, AR 11817 developed into a complex field configuration and produced a series of flares in the two days during August 11--12 (\tbl{tbl}). After the M1.5 flare on August 12, however, AR 11817 became dormant and it would wait until August 14 to produce C-class flares again, none of which, however, exceeded C5. 

\begin{figure}
  \centering
  \includegraphics[width=\hsize]{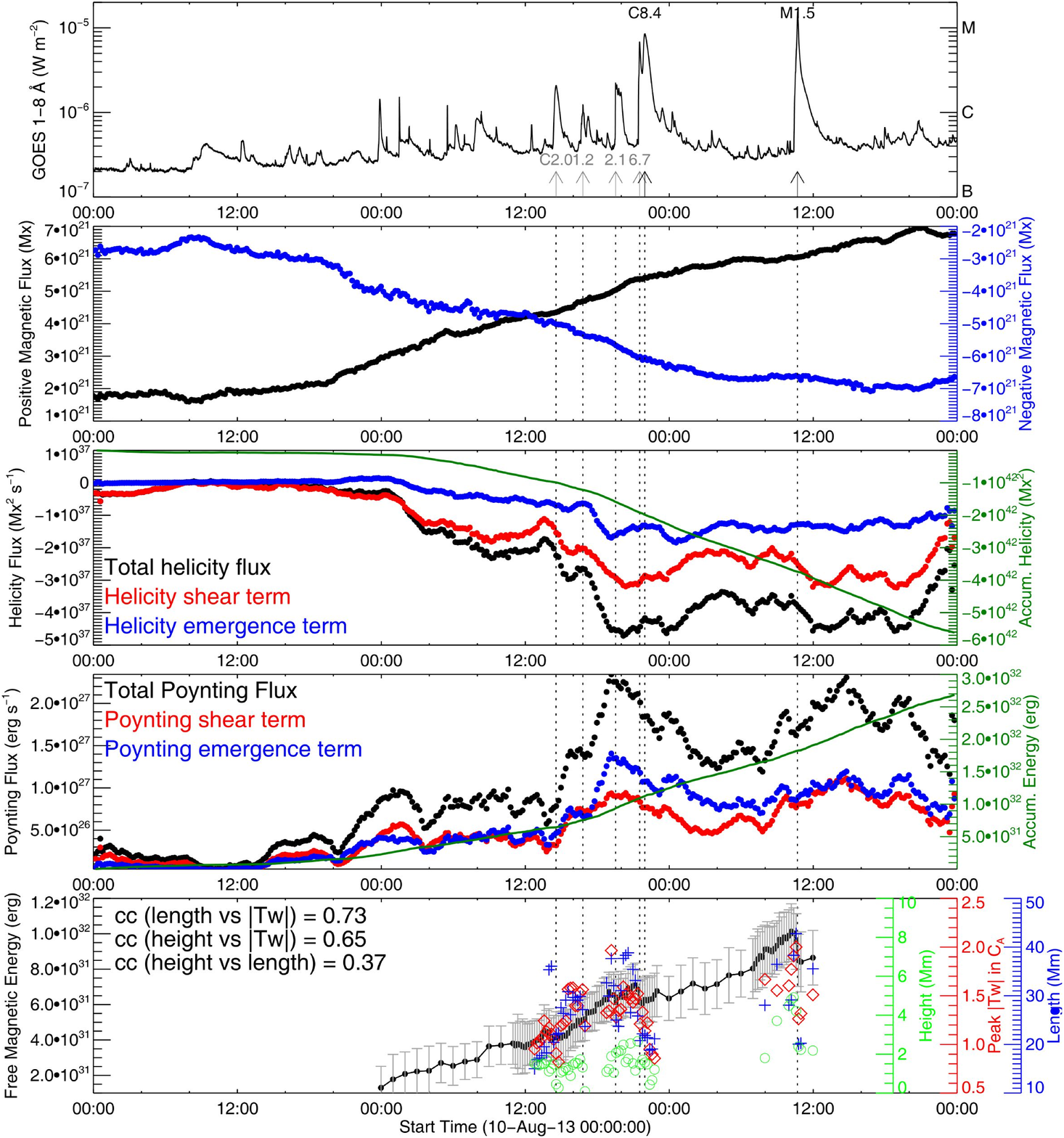}
  \caption{\small Helicity and energy injection into AR 11817 during August 10-12. The panels from top to bottom show the GOES 1--8~{\AA} light curve, the temporal profiles of magnetic fluxes, helicity fluxes, and Poynting fluxes, and the free magnetic energy derived from the NLFFF. The helicity and magnetic energy accumulated in the active region are displayed in green in the corresponding panel, scaled by the right y-axis. Vertical dotted lines indicate the peak times of the flares studied, whose GOES classes are annotated in the top panel, where confined/eruptive flares are marked by grey/black arrows, respectively. The peak $|\mathcal{T}_w|$ (red) within the MFR's cross section in the cutting plane $\mathrm{C_A}$ and the height(green) and length (blue) of the field line threading this peak $|\mathcal{T}_w|$ position at different times (see Figure~\ref{qfactor}) are shown in the bottom panel, scaled by the right y-axes. \label{helicity}}
\end{figure}

\tbl{tbl} lists all the C-and-above flares taking place during the three days 2013 August 10--12, during which the longitudinal center of AR 11817 changed from E48$^\circ$ to E10$^\circ$. The flares studied in this paper are numbered and boldfaced. Note that the C6.7 and C8.4 flares on August 11 are very close in time and practically can be regarded as a single flare with two peaks. However, the C6.7 flare is apparently confined, resulting in no CME, while for the subsequent C8.4 flare an ejection of a bubble-like structure leading up to a CME is seen in AIA images (not shown), which is hence conventionally classified as an eruptive flare. We ignored two C1 flares in this study: one occurred before any significant emergence of P2-N2; the other occurred during the decay phase of Flare No.~5 (see also the top panel of Figure~\ref{helicity}); both last for a duration of only several minutes, less than the cadence of the HMI vector magnetograms (12 min).

\begin{table}[h]
\caption{List of Flares} \label{tbl}%
\centering
\begin{tabu}{ccccccc} 
\toprule
No. & Date & Start & Peak & End & GOES Class & Type \\
\midrule
\nodata & 11-Aug-2013 & 05:26 & 05:30 & 05:32 & C1.1 & Confined \\
\rowfont{\bf} %
1 & 11-Aug-2013 & 14:18 & 14:34 & 14:49 & C2.0 & Confined \\
\rowfont{\bf} %
2 & 11-Aug-2013 & 16:37 & 16:48 & 16:55 & C1.2 & Confined \\
\rowfont{\bf} %
3 & 11-Aug-2013 & 19:27 & 19:32 & 20:03 & C2.1 & Confined \\
\rowfont{\bf} %
4 & 11-Aug-2013 & 21:24 & 21:31 & 21:38 & C6.7 & Confined \\
\rowfont{\bf} %
5 & 11-Aug-2013 & 21:47 & 21:58 & 22:11 & C8.4 & Eruptive \\
\nodata & 12-Aug-2013 & 00:10 & 00:15 & 00:20 & C1.0 & Confined \\    
\rowfont{\bf} %
6 & 12-Aug-2013 & 10:21 & 10:41 & 10:47 & M1.5 & Eruptive \\ 
\bottomrule
\end{tabu}
\end{table}

Figure~\ref{helicity} shows the helicity and energy injected into AR 11817 during August 10--12. The panels from top to bottom show the GOES 1--8~{\AA} light curve, the temporal profiles of magnetic fluxes, helicity and energy fluxes, and the free magnetic energy derived from the NLFFF. Significant flux emergence begins from the end of August 10. Correspondingly, the helicity injection rate is quickly enhanced, but the shear term is dominant over the emergence term. In contrast, the Poynting flux from the shear term is often slightly lower than that from the emergence term. This appears to be common among active regions \citep{liuy14}. The accumulated helicity reaches $-2\times10^{42}$ Mx$^2$ when the first eruptive flare (No.~5) took place on August 11, and $-4\times10^{42}$ Mx$^2$ when the second eruptive flare (No.~6) occurred on August 12. Both the helicity and Poynting fluxes decreased from about 20 UT on August 12 onward, which might explain that AR 11817 became flare-quiet after Flare No.~6.

\subsection{Structure and Stability of the MFR} \label{subsec:struct}
\begin{figure}
	\centering
	\includegraphics[width=0.8\hsize]{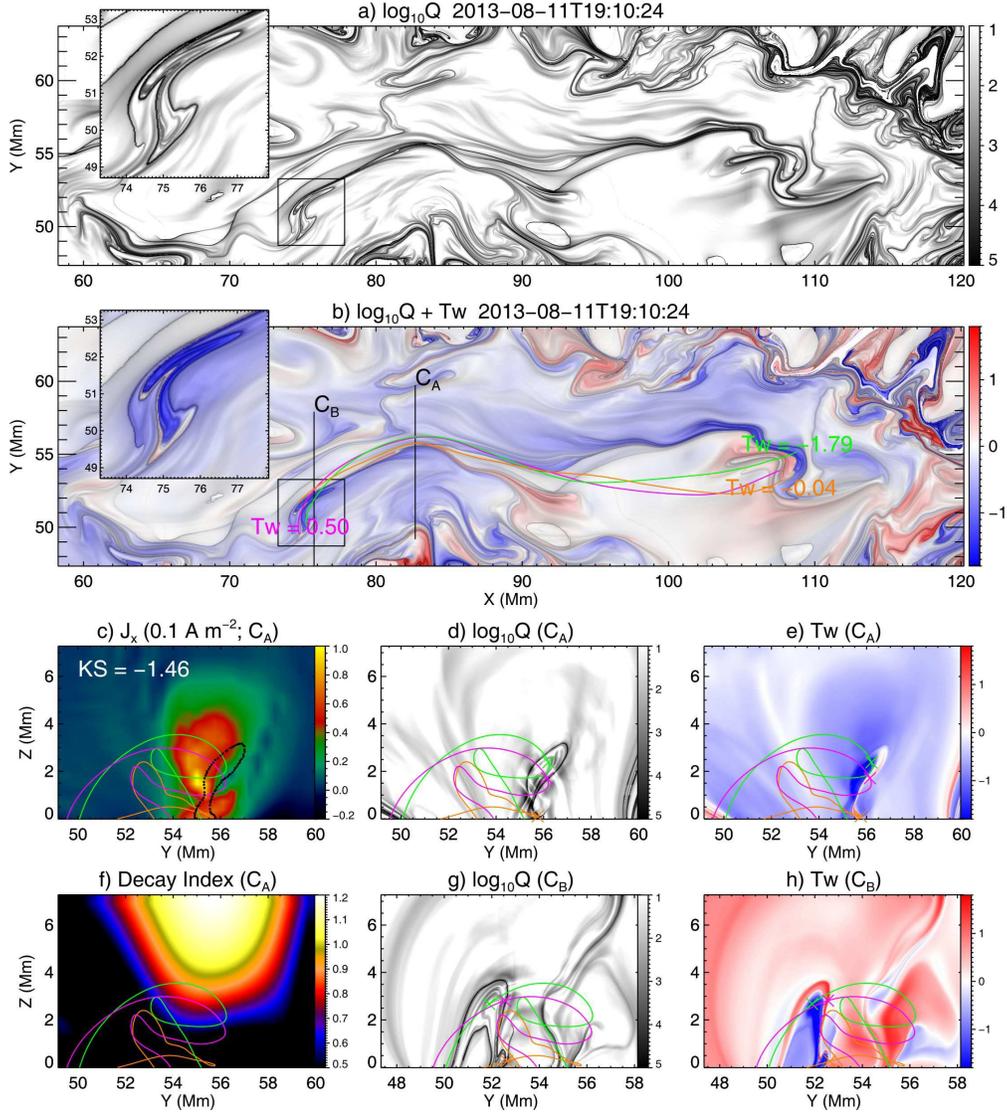}
	\caption{\small Various properties of AR 11817 before the C2.1 flare at 19:27 UT on 2013 August 11 (Flare No.~3; see also Figure~\ref{tw0811T18}). (a) Logarithmic $Q$ above 1 (white) and saturated at 5 (black). (b) Twist number saturated at $\pm1.8$ and blended with logarithmic $Q$. In (a) and (b) a rectangle encloses one of the MFR's footpoint regions. This rectangular region is enlarged and redisplayed in the upper left corner. (c--f) Current density $j_x$, logarithmic $Q$, twist number, and decay index in the cutting plane C$_A$ (denoted in (b)). (g--h) Logarithmic $Q$ and twist number in the cutting plane C$_B$ (denoted in (b)).  Representative field lines of the MFR (green), BPSS (orange) and QSL (magenta) are shown in both cutting planes. The crosses indicate where the field line threads the cutting plane. The cross-section of the MFR in C$_A$ is identified by clicking on its high $Q$ boundary, which is replotted as black dots in (c). An animation is available in the online version of the Journal, showing $\log_{10} Q$ and $\mathcal{T}_w$ at various cutting planes from $x=109.3$ Mm to $x=72.9$ Mm (viewed in the $-x$ direction).    \label{qfactor}}
\end{figure}

\begin{figure}
  \centering
  \includegraphics[width=\hsize]{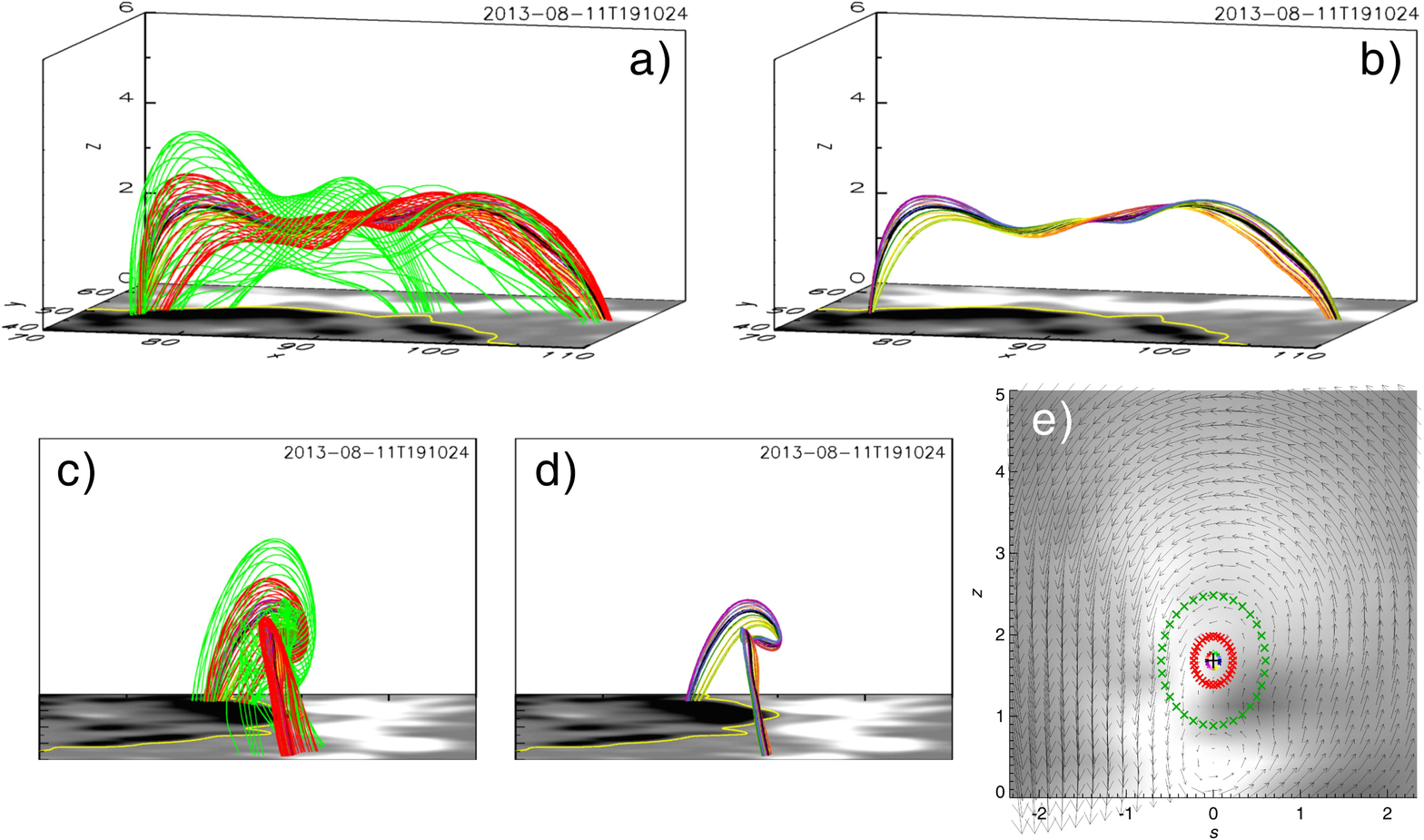}
    \caption{\small Three-dimentional perspective of the MFR indicated in Figure~\ref{qfactor}. The MFR axis is shown in black. Field line start points (marked by `x' symbols and in the same color as the field lines) are selected in the vertical cut plane (Panel (e); 77 deg to the $x$ axis), which is perpendicularly intersected by the MFR axis  at $(x,y,z)=(83.8,55.0,1.68)$ Mm (marked by a `+' symbol) and displays the normal current density component (gray scale) and in-plane field vectors. The horizontal axis denoted by $s$ is centered on the axis and in units of Mm. Panels (a) and (c) display all the field lines from two different perspectives, while Panels (b) and (d) show only the rainbow-colored field lines near the axis from the same perspective as (a) and (c), respectively. Physical units of the coordinates are Mm in (a), (b) and (e). \label{rope3d}}
\end{figure}

By the definition of MFRs (\S 1), the field lines winding around the rope axis must have similar magnetic connectivities, with the magnitude of their twist number exceeding unity. Hence, in the photospheric twist map (Figure~\ref{qfactor}(b)), taking Flare No.~3 for example, an MFR manifests itself straightforwardly as two conjugate compact regions with enhanced twist numbers of the same sign, which host the footpoints of the rope. The cross section of the MFR can be displayed by the twist number of field lines threading the cutting plane (Figure~\ref{qfactor}(e) and (h)). Comparing the photospheric (Figure~\ref{qfactor}(b)) and vertical (Figure~\ref{qfactor}(e) and (h)) twist maps with the corresponding $Q$-maps (Figure~\ref{qfactor}(a), (d), and (g)), one can see that the flux rope is bounded by high-$Q$ lines, i.e., a QSL that separates the twisted field from untwisted field \citep{titov02}. Several short bald-patch sections of the PIL are typically identified under the MFR in the magnetograms, indicating a complex topology that includes bald patch separatrix surfaces \citep[BPSSs;][]{titov&demoulin99}. In a bald patch the field lines graze the photosphere; such a representative field line is plotted in orange in Figure~\ref{qfactor}. Also shown are a representative twisted field line (green) and a QSL line (magenta). 

Figure~\ref{rope3d} shows perspective views of the MFR. The key to identifying and precisely locating the axis of the MFR is the vertical twist map. We have computed twist maps in many vertical cut planes, all oriented in the $y$ direction, in a wide range of $x$ values and traced a field line from the peak-$|\mathcal{T}_w|$ point in each map. It was found that all such field lines traced in the range $x=[74.5,107.3]$ Mm coincide within the limits of numerical accuracy. This includes nearly the whole axis of the MFR, except for a very short section ($\Delta x<2.5$ Mm) at its east end. Next we have placed a vertical cut plane at a point where the MFR axis runs horizontally (e.g., at the apex point) such that it intersects the cut plane perpendicularly and checked whether the in-plane field vectors display a rotational pattern centered at the intersection point, as expected for an MFR. This is indeed the case, and also the current density normal to the cut plane is enhanced in this area and thus consistent with the existence of an MFR (Figure~\ref{rope3d}(e)). Similarly to the $Q$ map in Figure~\ref{qfactor}(d), the MFR displays a rather compact and vertically elongated cross section. Choosing field line start points in the cut plane such that they follow this shape, the displays of Figure~\ref{rope3d}(a--d) are obtained. These confirm that a largely coherent MFR has formed in the range of strongest flux cancellation ($75\lesssim x\lesssim95$ Mm). The MFR extends further westward but is less twisted and less coherent in this area, where flux from various photospheric flux patches joins the forming rope.

This analysis was also performed for the NLFFF before the Flares No.~4 (Aug~11, 21:22~UT) and No.~6 (Aug~12, 10:22~UT), confirming that the peak-$|\mathcal{T}_w|$ point in vertical twist maps locates the axis of the MFR in the considered AR. Some care must be executed, however, as $\mathcal{T}_w$ may peak away from the MFR axis in other configurations, where the current has a broad spatial distribution (uniform across the rope or even a hollow profile). In such cases, $\mathcal{T}_w$ has a minimum at the MFR axis (which also facilitates locating the axis; see Appendix~\ref{append:tw}). Moreover, in areas where the MFR is not fully coherent it may not be possible to define an axis and $\mathcal{T}_w$ may simply peak at a long field line that runs through a volume of high shear.

The well-defined boundary of the main flux rope gives us an opportunity to test its stability with respect to the helical kink mode against the classical theory, namely, the Kruskal-Shafranov limit of one field line turn about the MFR axis, modified for the presence of photospheric line tying, which raises it to $N=1.25$ or $\Phi=2.5\pi$ for the twist angle $\Phi(r)=B_\phi L_z/(rB_z)$. Using $B_\phi=\mu_0I_z/(2\pi r)$, a limiting axial current $I_\zeta$ for an idealized cylindrical flux rope with radius $r_0$ and length $L_z$ is obtained \citep{Schindler2006}, 
\begin{equation}
I_\zeta=1.25\frac{(2\pi r_0)^2B_z(r_0)}{\mu_0L_z}\,. \label{eq:Izeta}
\end{equation} 
Based on this, we introduced a dimensionless parameter,
\begin{equation}
\text{KS} \equiv \frac{F_p}{F_t}=\frac{\int_0^{r_0}\int_0^{L_z}B_\phi(r) \,dr\,dz}{\int_0^{r_0}\int_0^{2\pi}B_z(r) r\,dr\,d\phi}\,,
\end{equation} where $F_p$ is the poloidal flux and $F_t$ the toroidal flux. One can immediately see that KS is related to N with two assumptions (simplicities): 1) $B_\phi$ is distributed linearly over the radius from the center of the rope, or equivalently, $I_z$ has a uniform distribution in the cross-section of the rope, and 2) $B_z$ is uniformly distributed across the cross section, i.e.,
\begin{align}
\text{KS} & \approx \frac{B_{\phi0}r_0L_z/2}{B_{z0}\pi r_0^2}=N(r_0) \nonumber\\
		  & = \frac{\mu_0 I_z L_z}{(2\pi r_0)^2 B_{z0}}\,,\nonumber
\end{align} 
where $B_{\phi0}$ and $B_{z0}$ are given at $r=r_0$. With Eq.~\ref{eq:Izeta}, one can see that the threshold of $|\mbox{KS}|=1.25$. In our case, we integrated $j_x$ over the rope's cross section in the cutting plane C$_\text{A}$ (Figure~\ref{qfactor}) to estimate $F_p$, i.e., $F_p = \mu_0L/4\pi\int j_x\,dS$, where $L\approx 37.83$~Mm is given by the length of the rope axis in Figure~\ref{rope3d}. Similarly, $F_t$ is estimated by integrating $B_x$ over the same cross section. The result gives $\text{KS} = -1.46$, exceeding the KS threshold. In comparison, for a group of selected field lines near the rope axis (rainbow colored in Figure~\ref{rope3d}), we found that $<T_w>=-1.88\pm0.06$ with $\mathrm{max}(T_w) = -1.96$ and $<T_g>=-1.28\pm0.20$ with $\mathrm{max}(T_g) = -1.49$. Hence, both the integral property of the rope as characterized by KS and the peak ``local'' twist number ($\mathcal{T}_w$ or $\mathcal{T}_g$) of individual field lines (near the flux rope axis) suggest that the MFR may be marginally kink-unstable. However, the exact threshold of instability depends on further parameters, in particular on the strength of the external toroidal (shear) field component \citep{Kliem&al2014}, and it is also higher for very slim flux ropes \citep{Baty2001} and if the rope is not fully coherent, as in the present case. Moreover, MFRs do not always have a well-defined high-Q boundary \citep[see, e.g.,][]{liu14}; for the current MFR, the eastern footpoint region is fully enclosed by high-Q lines, but the western footpoint region is only partially enclosed (see Figure~\ref{qfactor}(b)). Hence, the above analysis cannot be easily applied to all occasions.

Therefore, we tested the stability of the MFR by using the NLFFF as
the initial condition in an MHD code, employing the same grid. We used a simple density model which yields a slow decrease of the Alfv\'en velocity with distance from the flux concentrations \cite[see, e.g.,][]{torok&kliem05}. No perturbation was applied. Only relatively small, nearly vertical motions of the MFR ensued spontaneously, which subsequently decreased monotonically. The resulting stable numerical MHD equilibrium shows an MFR close to the one in the corresponding initial NLFFF. Similar results are obtained for the MFR before Flares No.4 (21:22~UT on August 11) and 6 (10:22~UT on August 12). We conjecture that the stability of the MFR in the NLFFF only shortly before or just at the onset of the flares results from the minimum-current assumption in the disambiguation of the vector magnetograms.

Further, we calculated in the cutting plane C$_A$ the decay index $n=-d\ln B_t/d\ln h$, where $h$ is the height above the photosphere and $B_t$ is the transverse component of the potential field.  Numerical experiments have demonstrated that an MFR becomes torus unstable when the decay index of the external poloidal field reaches a critical value of $\sim\,$1.5 \citep[e.g.,][]{Torok2007a}, and $B_t$ is a good proxy of the external poloidal field at the axis of the MFR \cite[e.g.,][]{Kliem&al2013}. The twist and $Q$ maps in C$_A$ clearly show that the MFR is located below about 4 Mm above the photosphere (Figure~\ref{qfactor}(e) and (h)). In comparison with the map of the decay index (Figure~\ref{qfactor}(f)), one sees that the MFR resides in a region stable to the torus instability unless it experiences a large disturbance that pushes it up to much higher altitudes. In fact, for all the flares studied, the MFR is located below the $n=1$ contour (see the insets in Figures~\ref{tw0811T14}--\ref{tw0812T10}). In cases No.~4 (confined) and 6 (eruptive), the MFR is relatively high-lying, reaching the $n=1$ contour before the flare (Figures~\ref{tw0811T21} and \ref{tw0812T10}), but still well below the threshold of the torus instability ($n\sim1.5$). 

Hence, throughout the time series analyzed, it appears that the MFR is close to the threshold of the helical kink instability but far away from that of the torus instability. We propose a scenario in which the helical kink mode commences at the onset of all flares studied and the consequent lifting of the MFR does not reach the threshold height for the onset of the torus instability in the confined Flares No.~1--3, but, starting from a greater height, reaches it in the eruptive Flare No.~6 as well as the combined eruption seen as Flares No.~4 and 5. 

\subsection{Indication for the Existence of a Double-decker Flux Rope} \label{subsec:dbl}

\begin{figure}
	\centering
	\includegraphics[width=\hsize]{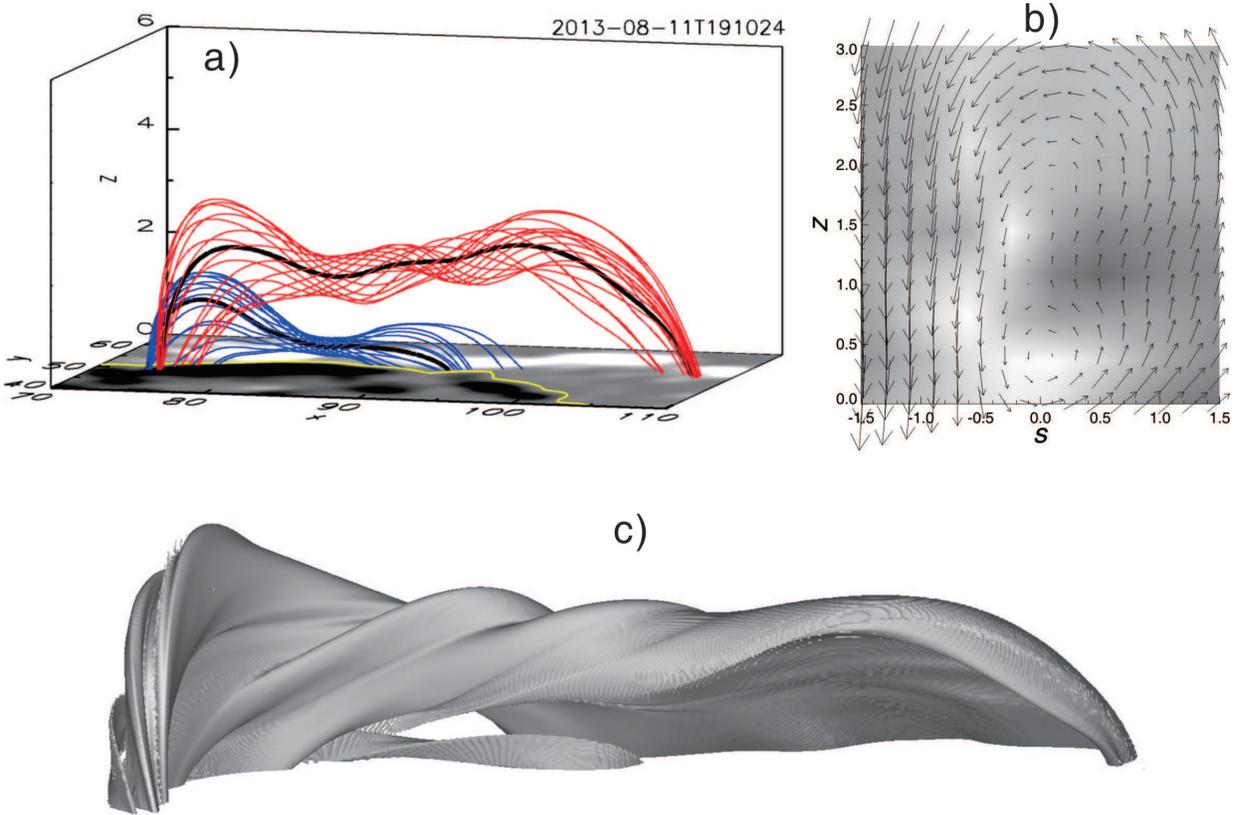}
	\caption{\small  A double-decker MFR configuration.a) Twisted Field lines of the double-decker MFR in the NLFFF on 2013 August~11 at 19:10~UT, in the same format as Figure~\ref{rope3d}. b) Normal current density and in-plane field vectors in a vertical cut plane oriented perpendicularly to the MFR axis (77 deg to the $x$ axis) at $(x,y)=(83.9,54.9)$ Mm. c) Isosurface of $T_w\leq -1$ viewed from the same perspective as Panel (a). Physical units of the coordinates are Mm in (a) and (b).\label{dblfr}}
\end{figure}

The vertical twist map in Figure~\ref{qfactor}(e) reveals an unexpected layer of reversed twist (and $\alpha$) on the north edge of the MFR and an area of slightly enhanced standard (negative) twist values under the MFR and the layer. The nearby vertical cut in Figure~\ref{rope3d}(e) shows enhanced current density and a rotational pattern of the in-plane field components in this area, i.e., an indication for the existence of a second, smaller MFR under the main MFR. Such a double-decker flux rope configuration was suggested previously as a candidate for partial eruptions \citep{liu12, Cheng2014, Zhu&Alexander2014}. We have investigated the structure using our series of vertical twist maps, also rotating the cut planes about the vertical, to better display the axial current and azimuthal field structure (Figure~\ref{dblfr}(b)). A shallow peak of the twist number, $\mathcal{T}_w\approx-1.1$, is found, which coincides with a shallow maximum of the current density and the center of a rotational structure of the azimuthal field component (Figure~\ref{dblfr}(b)). The rotational structure exists all around the center point, clearly indicating that the region is a flux rope, not simply a sheared volume, and that the peak-$|\mathcal{T}_w|$ point lies on the MFR axis. This is substantiated by the field line plot in Figure~\ref{dblfr}(a) which displays a double-decker configuration of two left-handed flux ropes, and by the isosurface of $\mathcal{T}_w \leqslant -1$ (Figure~\ref{dblfr}(c)).

A shear layer is a natural consequence if both flux ropes in a double-decker configuration have the same handedness. The field lines in the bottom layer of the upper rope then have a direction different from the field lines in the top layer of the lower rope \citep[see Figure 12 in][]{liu12}. If the system is driven, such that the ropes evolve in flux content and position, currents are likely induced in the shear layer, i.e., $\alpha\ne0$ and hence $\mathcal{T}_w\ne0$. If the direction of the axial field in the configuration is uniform, like in the present case, then the axial current, $\alpha$, and $\mathcal{T}_w$ in the shear layer are opposite in sign to the two MFR, in agreement with Figure~\ref{qfactor}(e). Since oppositely directed currents repel each other, the shear layer should have a stabilizing influence on the configuration. 

\subsection{Change of Magnetic Twist Across Flares}
\begin{figure}
  \centering
  \includegraphics[width=0.9\hsize]{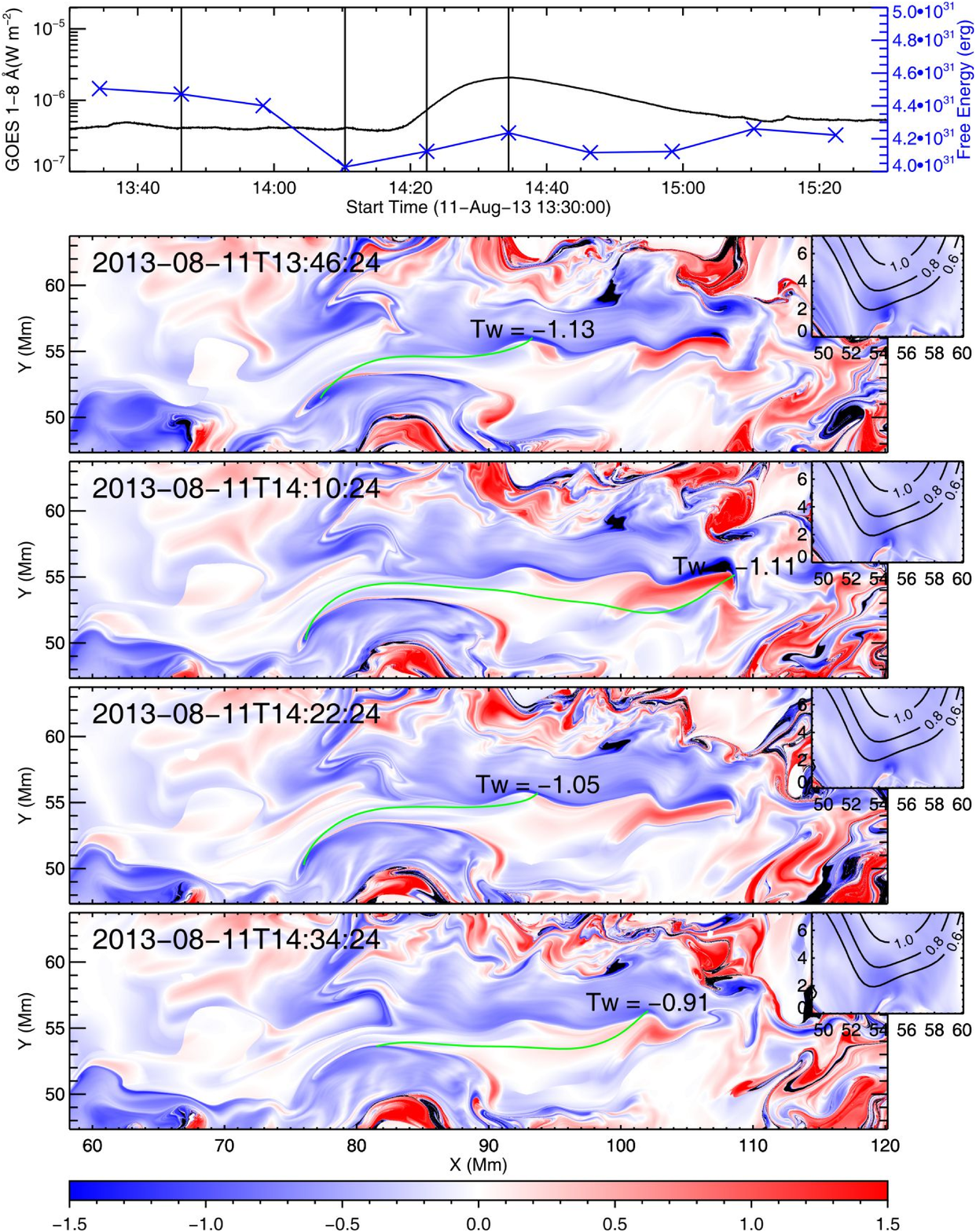}
  \caption{\small Twist maps before and after the C2.0 (No.~1) flare on 2013 August 11. The top panel shows the GOES 1--8~{\AA} light curve, scaled by the left y-axis, and the free magnetic energy derived from NLFFF modeling, scaled by the right y-axis. The vertical lines in the top panel mark the times of the vector magnetograms, based on which the twist maps in the panels below are calculated. Superimposed on the twist maps are representative field lines, whose twist numbers are annotated at the corresponding footpoints. Saturated blue colors are shown as black. The insets present the twist maps in the same cutting plane C$_A$ (at $X= 82.7$ Mm) as in Figure~\ref{qfactor}, and are superimposed by contours of the decay index. \label{tw0811T14}}
\end{figure}

\begin{figure}
  \centering
  \includegraphics[width=0.9\hsize]{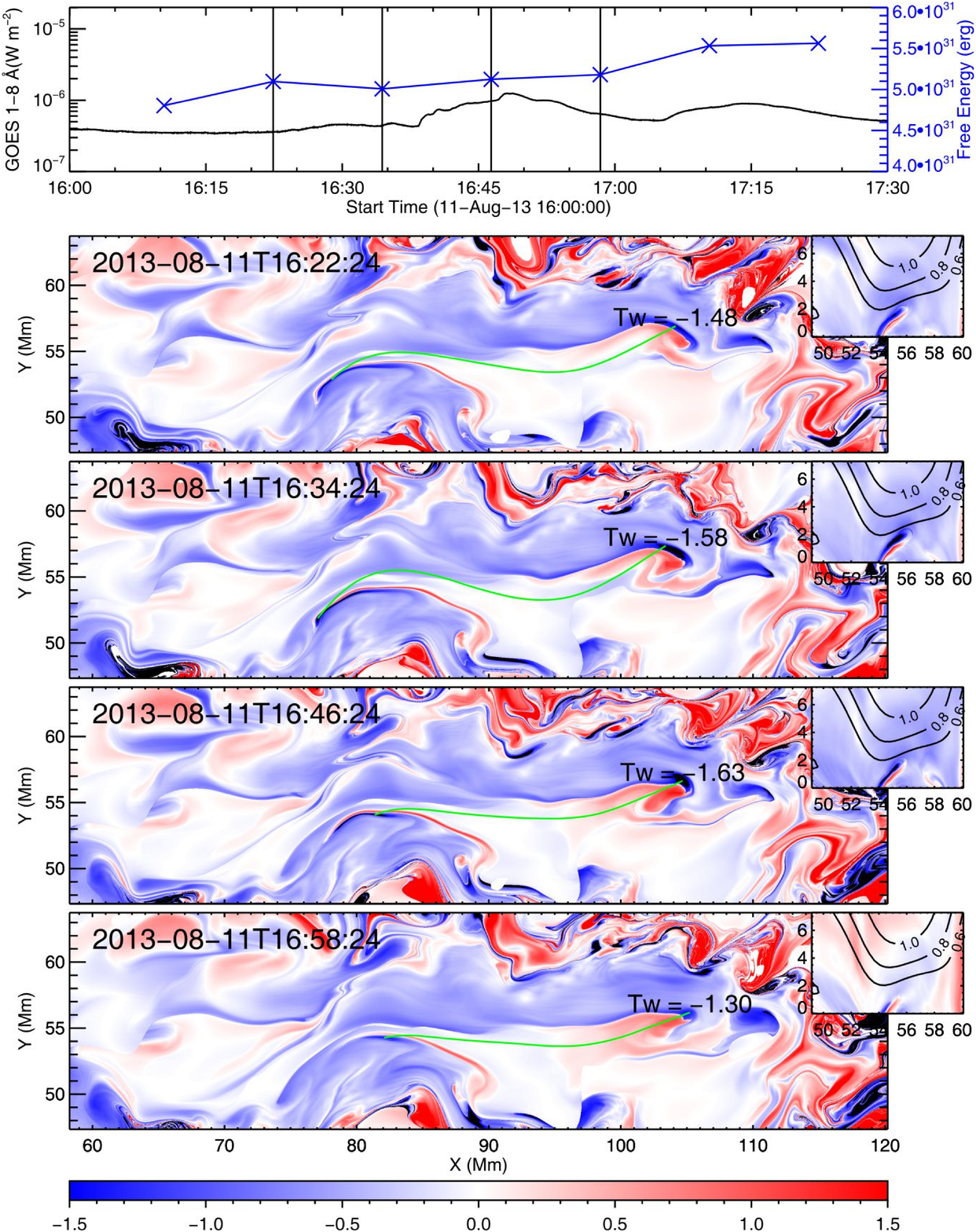}
  \caption{\small Twist maps before and after the C1.2 flare (No.~2) on 2013 August 11. Same layout as Figure~\ref{tw0811T14}. \label{tw0811T16}}
\end{figure}

\begin{figure}
  \centering
  \includegraphics[width=0.9\hsize]{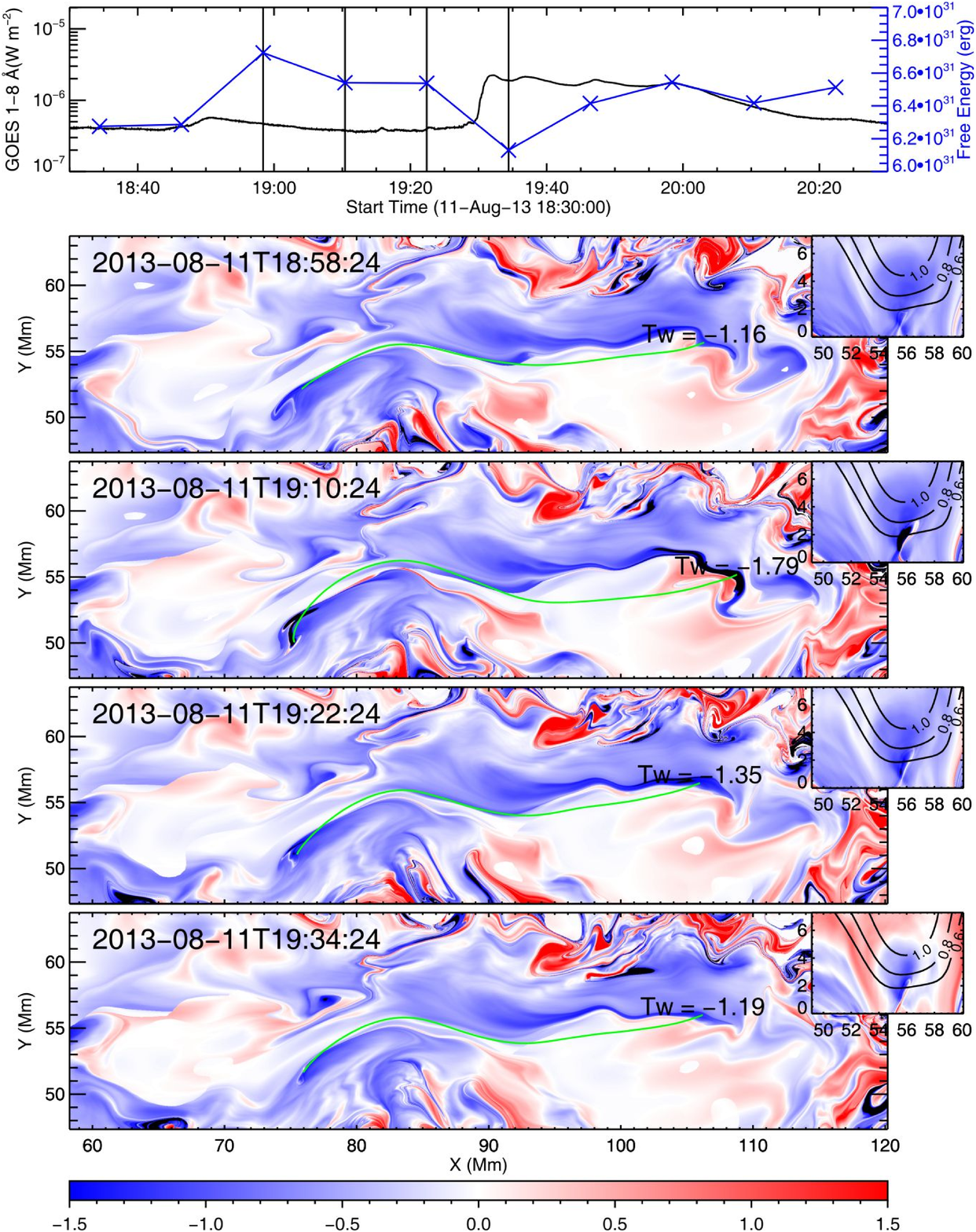}
  \caption{\small Twist maps before and after the C2.1 flare (No.~3) on 2013 August 11. Same layout as Figure~\ref{tw0811T14}.\label{tw0811T18}}
\end{figure}

\begin{figure}
  \centering
  \includegraphics[width=0.7\hsize]{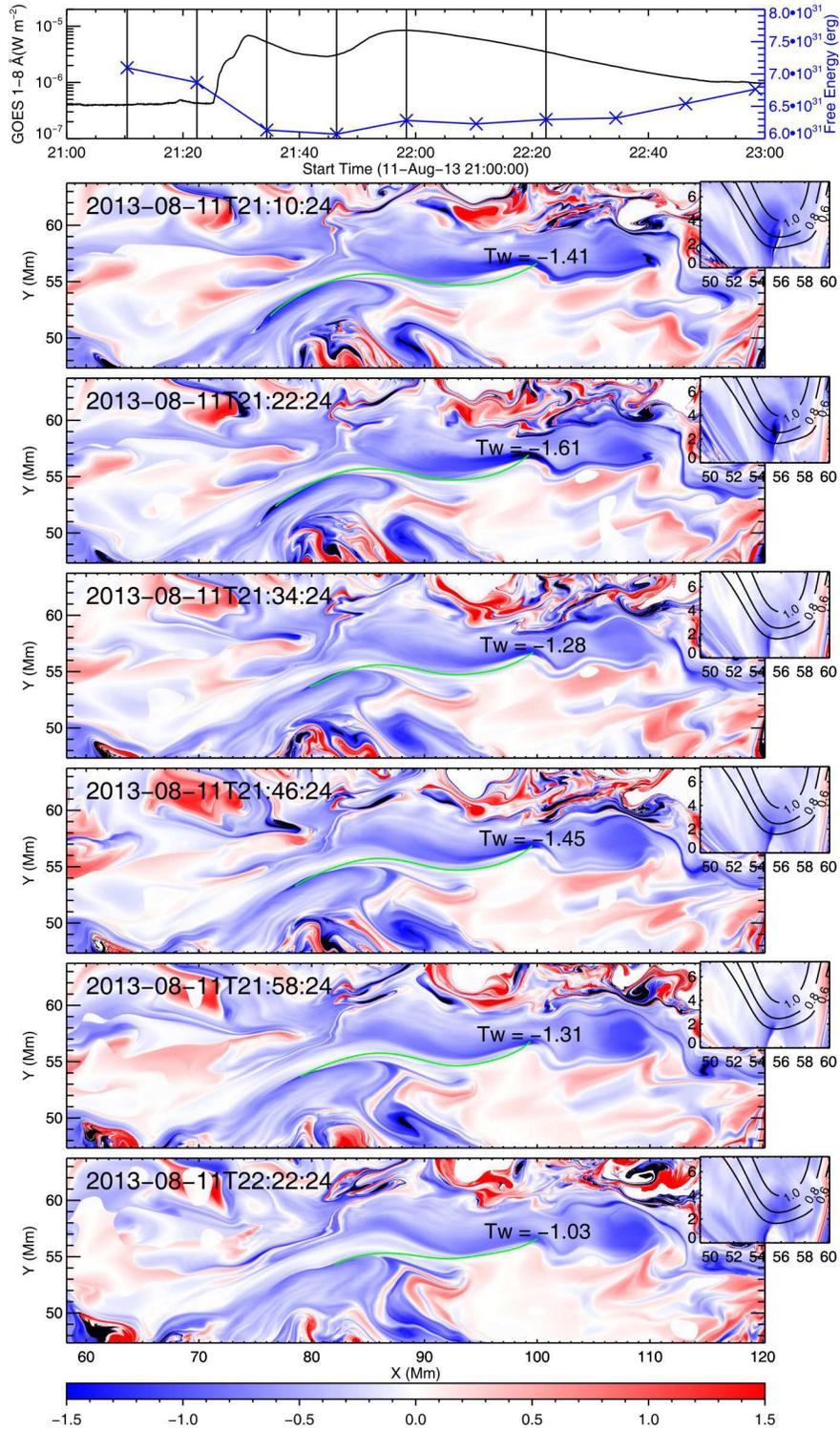}
  \caption{\small Twist maps before and after the C6.7 (No.~4) and C8.4 (No.~5) flare on 2013 August 11. Same layout as Figure~\ref{tw0811T14}. \label{tw0811T21}}
\end{figure}

\begin{figure}
  \centering
  \includegraphics[width=0.9\hsize]{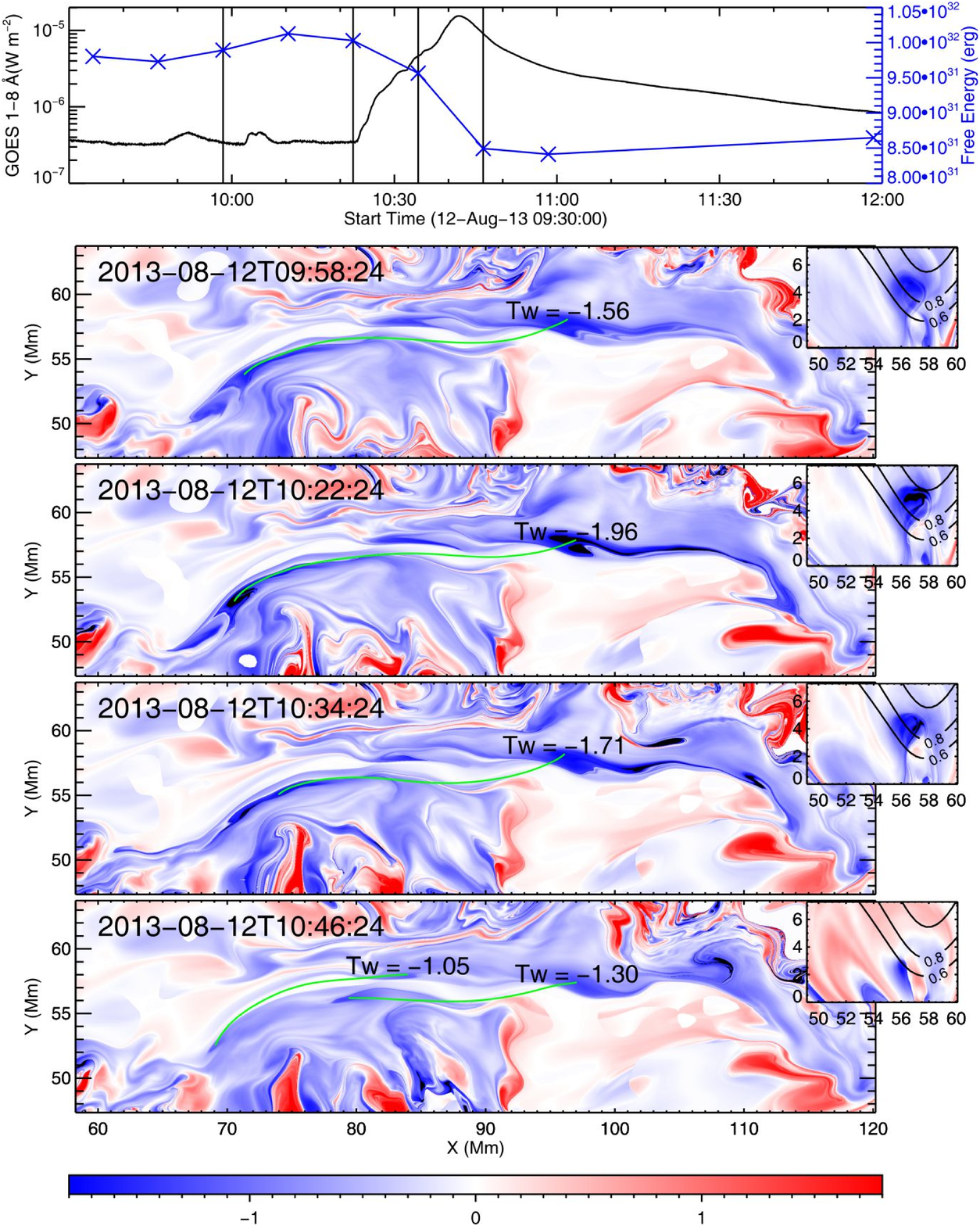}
  \caption{\small Twist maps before and after the M1.5 flare (No.~6) on 2013 August 12. Same layout as Figure~\ref{tw0811T14}, except that the scale of the color bar is expanded to $\pm1.8$ for better visualization of the twist maps. \label{tw0812T10}}
\end{figure}

We calculated a series of photospheric twist maps based on the NLFFF modeling, focusing on a small region (black rectangle in Figure~\ref{bfield}(f)) that encloses the major PIL, where the magnetic field is highly sheared and most shearing and converging flows are detected. Signatures of an MFR along the major PIL start to appear in the twist map from about 12:46 UT on August 11, about 1.5 hours before Flare No.~1, and become quite prominent before the occurrence of the C2.1 flare (No.~3) at 19:27 UT on August 11. One can see from Figure~\ref{qfactor} that both the footpoint regions and the cross sections of the MFR are featured by dark blue colors, in agreement with the negative helicity dominating in this active region, and enclosed by high-$Q$ lines (Figure~\ref{qfactor}), suggesting the presence of a QSL at the rope boundary. If a coherent MFR appears in the twist map calculated in a cutting plane by visual inspection, we recorded the peak $|\mathcal{T}_w|$ within the MFR's cross section. The result is plotted in the bottom panel of Figure~\ref{helicity}, which shows a common trend that the peak $|\mathcal{T}_w|$ temporarily increases before the flare while it decreases after the flare peak, both for confined and eruptive flares. This is also seen in the environment of the peak-$|\mathcal{T}_w|$ point. We will now examine this trend case by case.

Figures~\ref{tw0811T14}--\ref{tw0812T10} show the twist maps across each flare studied (Table \ref{tbl}). The decrease in magnetic twist is often related to a decrease in the free magnetic energy of the overall active region (except Flares No.~2 and 5), but due to the relatively small volume of the MFR, the transient increase in magnetic twist does not translate to a similar increase in the free magnetic energy, although the latter possesses a gradually increasing profile with time (Figure~\ref{helicity}). It is remarkable that, despite that Flares No.~4 and 5 are lumped together, the twist map still indicates a temporary increase in magnetic twist just prior to Flare No.~5. In comparison, although it clearly decreases across the combined eruption seen as Flares No.~4 and 5 (Figure~\ref{tw0811T21}), the free magnetic energy does not change appreciably across No.~5. In Flare No.~2 (Figure~\ref{tw0811T16}), the weakest of our sample, the flare-related energy release is too weak to stand out against the general trend of energy build-up in the active region, but the change of magnetic twist across this flare is visible. For Flare No.~1 (Figure~\ref{tw0811T14}), the free magnetic energy decreases already 10--20~min before the flare \citep[see also][]{jing09}. A similar issue happens to the magnetic twist when its change is detected 10--20~min before Flare No.~3 (Figure~\ref{tw0811T18}). This might indicate a limited accuracy of the underlying magnetograms or of the NLFFF model, but it could also indicate that the flare is triggered by a photospheric evolution that starts already before the flare. We also noted that the height of the MFR gradually increases with time during the sequence studied (compare the insets in Figures~\ref{tw0811T14}--\ref{tw0812T10}).

We assume that the rise of the peak twist number $|\mathcal{T}_w|$ at the axis of the MFR just prior to the flares is correlated with the on-axis value of the classical twist number $|\mathcal{T}_g|$. The relationship between these twist numbers is analyzed in Appendix~\ref{append:tw}, where we find that, at the axis, $\mathcal{T}_w$ approaches the sum of $\mathcal{T}_g$ and a geometry-dependent term which measures the deviation from cylindrical symmetry (see Eq.~\ref{eq:Tg&Tw}). The expectation that such a deviation is of minor influence on the stability of the MFR compared to the twist underlies our assumption, which should be substantiated in future investigations.

The peak magnetic twist at the axis of the MFR can increase due to a twisting of the rope by rotational motions at its footpoints, or due to a lengthening of the rope by reconnection. The latter involves a change in the footpoints and likely also leads to a greater height of the rope. In the derived velocity maps by DAVE4VM, there are no obvious counterclockwise rotations associated with the footpoint regions as expected for the buildup of the left-handed twist, but we did find positive correlations among the peak $|\mathcal{T}_w|$, its height, and the length of the field line threading the peak $|\mathcal{T}_w|$ position (bottom panel of Figure~\ref{helicity}). Thus, primarily reconnection should have caused the transient increase in $|\mathcal{T}_w|$ prior to the flares.

For eruptive flares, it is readily understood that the decrease in magnetic twist must result from the expulsion of the flux rope, although a flux rope still survives the eruption in both cases studied (Flares No.~5 and 6), suggesting that most likely a partial expulsion of the flux rope occurs during the eruption. In Flare No.~5, the remnant MFR is significantly reduced in flux, height, and twist (Figure~\ref{tw0811T21}), eventually disappears, and then reforms about 2 hrs before Flare No.~6 (see the bottom panel of Figure~\ref{helicity}). 
In that flare, the MFR splits into two remaining branches at a lower height immediately after the flare peak (Figure~\ref{tw0812T10}), suggesting that a reconfiguration of the flux rope might be ongoing. It is interesting that the two branches take the form of two interlocking Js, which is the reverse of the double-J-to-S transformation that is typical of tether-cutting reconnection \cite[e.g.,][]{liu10}. For confined flares, the decrease in magnetic twist must indicate a conversion of magnetic energy primarily into plasma heating, but what mechanism causes this decrease is unclear until we have further analyzed Flare No.~3 using the high-resolution NST data (\S\ref{subsec-kink}).

\subsection{C2.1 Flare on 2013 August 11} \label{subsec-kink}
\begin{figure}
	\centering
	\includegraphics[width=0.8\hsize]{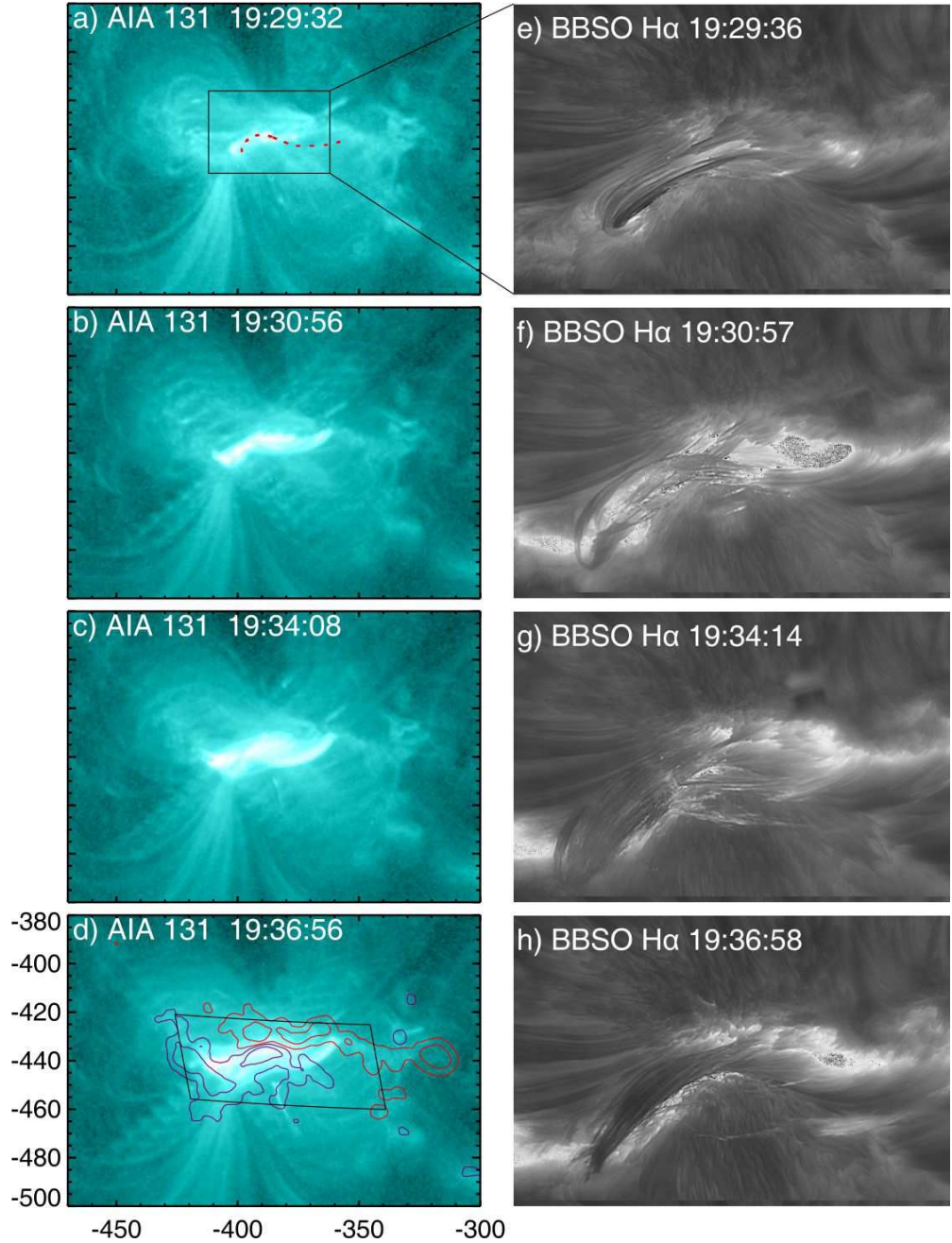}
	\caption{\small Snapshots of Flare No.~3 observed in 131~{\AA} by the space-borne AIA (top) and in H$\alpha$ by the ground-based NST. The rectangle in (a) indicates the FOV of the NST observation. The warped rectangle in (d) shows the FOV of the calculated photospheric Q-maps and twist maps (Figure~\ref{qfactor} and Figure~\ref{tw0811T18}). The contours denote the magnitude of the local $B_r$ at the levels of $\pm200$ and $\pm800$ G, with red (blue) indicating positive (negative) polarities. \label{halpha}}
\end{figure}

\begin{figure}
	\centering
	\includegraphics[width=\hsize]{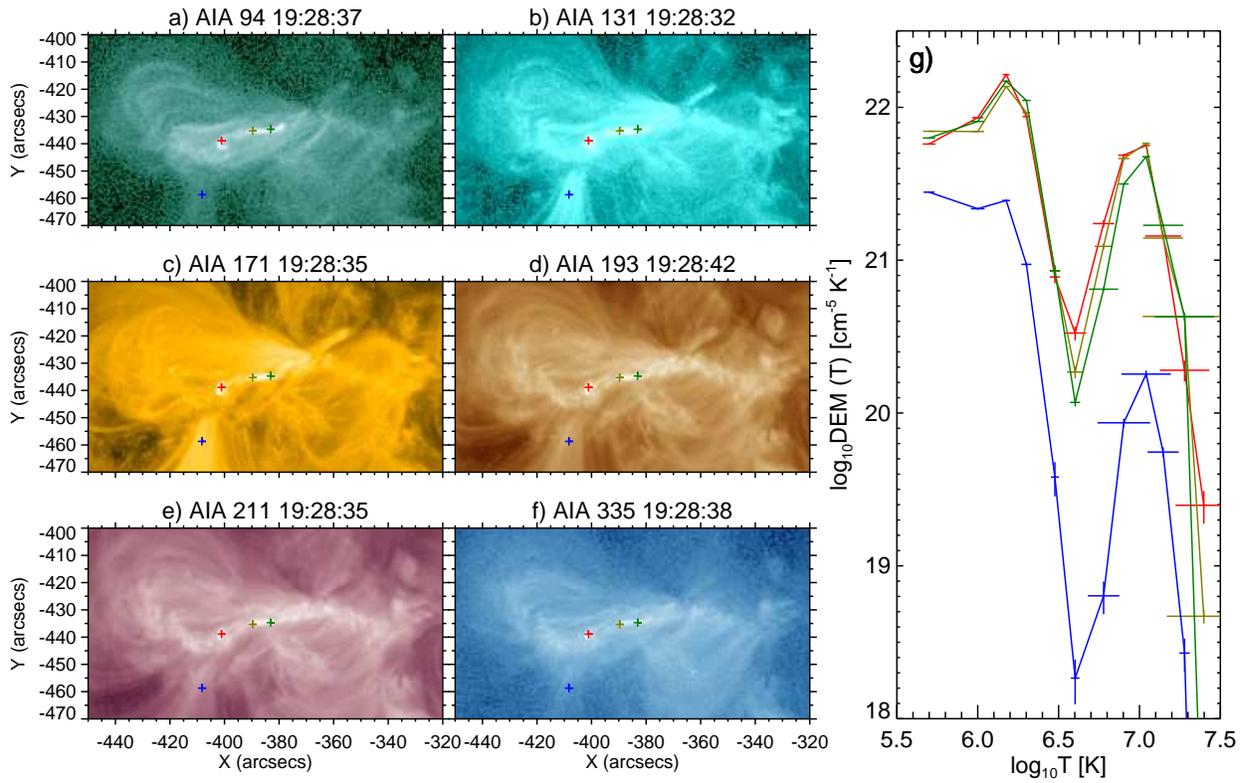}
	\caption{\small DEM analysis of the sigmoid at the onset of Flare No.~3. Panels (a--f) show the sigmoid as seen in the six EUV channels of AIA. Panel (g) displays the DEMs for selected points (color coded).   \label{hot}}  
\end{figure}

Here we highlight the confined C2.1 flare taking place at 19:27 UT on August 11 (Flare No.~3). With the aid of NST observations \citep[see the detailed analysis by][]{wang15}, this event sheds some light on the flaring mechanism. Figure~\ref{halpha} shows the flaring process observed in AIA 131~{\AA}, which manifests as the brightening of a small sigmoid extending less than 50$''$ in the east-west direction. In the high-resolution H$\alpha$ images, however, one sees a filament with a similar reverse S shape. \citet{wang15} have demonstrated that the filament traces the MFR. The filament axis is apparently writhed after the flare onset (Figure~\ref{halpha}(f--g)), with the counterclockwise rotation being consistent with the conversion of left-handed twist \citep{green07}. The deformation of the filament axis appears to be moderate and the filament soon settles back down in approximately six minutes to a similar configuration as before the flare, indicating that the flare is confined. The fine threads of the filament do not appear to be twisted by significantly more than one turn, which is consistent with the range of twist values of the MFR derived from the NLFFF modeling (see Appendix~\ref{append:tw}). Superimposing on the AIA 131~{\AA} image a representative field line of the MFR in the pre-eruption NLFFF (red dotted line in Figure~\ref{halpha}), one can see that the MFR matches the sigmoid very well. This is consistent with the sigmoid being energized beneath the erupting MFR \cite[e.g.,][]{titov&demoulin99}. 
In the process of eruption the coherence of the filament is distorted, with some brightenings occurring in its rising part (Figure~\ref{halpha}(f--g)). This suggests that reconnection also occurs with the overlying field, which halts the eruption according to the analysis of the decay index in \S\ref{subsec:struct}. Thus, although the writhe must be back-converted into twist as the filament settles back near its original shape, the MFR loses part of its twist by reconnection with the overlying field.

A diagnosis of the flaring plasma with the DEM method can shed more light on the energy release mechanism of this flare. One can see that immediately after the flare onset, the sigmoid is visible in each of the six EUV passbands of AIA (Figure~\ref{hot}). We obtain the average DEMs for 3 by 3 pixels centered on three selected loci along the sigmoid (red, olive and green crosses) as well as on neighboring coronal loops (blue cross) to serve as a reference (Figure~\ref{hot}(e)). The DEMs of the sigmoid  exhibit a significant peak at about 10 MK, besides the peak at about 1 MK that can mainly be attributed to the coronal background. Both the cool and hot DEM peaks of the sigmoid are significantly elevated above those of the neighboring coronal loops, suggesting that the sigmoid is both dense and hot. This sigmoid is obviously associated with the erupting filament observed in H$\alpha$; however, its size and shape do not follow the expanding and writhing filament, but rather remain close to the initial state (\fig{halpha}). This corresponds to the standard picture of the sigmoid being formed by current dissipation in a separatrix or quasi-separatrix layer underneath an activated flux rope. Figure~\ref{qfactor}(d) indeed shows the highest squashing factor where the QSL intersects itself under the MFR; this intersection is suggestive of the presence, or beginning formation, of a hyperbolic flux tube, where the flare reconnection is expected to develop \cite[e.g.,][]{Savcheva&al2012}. A location of the sigmoid under the flux rope is also supported by its high density. 

\section{Summary and Conclusions} \label{sec:conclusion}
Here we summarize what we have learned from the observations of a sequence of four confined and two subsequent eruptive flares and the associated vector magnetograms in AR~11817, and how we interpret these observations.

\begin{itemize}
\item An MFR exists in the NLFFF extrapolated from the magnetograms prior to each flare in the series (where the closely associated Flares No.~4 and 5 are considered to be a compound event). Signatures of the MFR start to appear about 1.5 hours before the first confined flare, and it is reformed after a partial expulsion in the first eruptive flare.

\item The MFR's peak twist number $|\mathcal{T}_w|$ temporarily increases within half an hour before each flare, while it decreases after the flare peak, for both confined and eruptive flares. The temporary increase in $|\mathcal{T}_w|$, assumed to reflect an increase in magnetic twist about the MFR axis, $|\mathcal{T}_g|$ (see Appendix~\ref{append:tw}), can only partly be attributed to the photospheric helicity injection, which is more or less quasi-static in nature, but signifies the additional importance of reconnection, in the present case including the axis of the MFR as well as tether-cutting reconnection in the outer parts of the rope, at the photospheric/choromospheric level. Both are expected to occur episodically as opposite-polarity magnetic elements are randomly brought into contact at the PIL by photospheric flows. The reconnection heats rope-like structures in the corona (termed `hot channels' by some authors) shortly before flares \citep[e.g.,][]{liu10,zhang12}. For confined flares, two interrelated mechanisms could be responsible for the reduction in the MFR's twist: 1) onset of the helical kink instability, which converts twist into writhe; 2) current dissipation at the MFR's boundary, i.e., QSL, which reconnects twisted field with the surrounding, untwisted field, so as to distribute magnetic twist into a larger volume. The latter is expected to be most efficient when the MFR loses stability and writhes. For eruptive flares, the reduction of magnetic twist in the source region is due to the (partial) expulsion of the MFR.

\item The temporary increase in the MFR's twist prior to flares leaves little imprint in the AR's free magnetic energy, while the decrease in the MFR's twist after flares corresponds to a stepwise decrease in free magnetic energy in most of the flares studied. This suggests that the additional free magnetic energy that is related to the
trigger of the eruptions is very localized and modest compared with that released during the eruptions. This feature might have important implication for space weather forecasting. In contrast, none of the parameters derived for the whole active region, including magnetic flux, helicity and energy, gives such a clear signal for impending eruptions \citep[see also][]{leka&barnes07}.

\item The MFR of interest is moderately twisted (peak $|\mathcal{T}_w| \lesssim 2$ and peak $|\mathcal{T}_g| \lesssim1.5$). This is derived from the local twist number $\mathcal{T}_w$, the general definition of twist number $\mathcal{T}_g$, and the global Kruskal-Shafranov criterion in our series of NLFFF models and indicates that the MFR is close to the threshold of the helical kink instability. The high-resolution NST observations of Flare No.~3 also indicate the onset of this instability. A moderately twisted MFR results in mild writhing of the filament axis when the helical kink mode sets in, which can only be discerned with a spatial resolution much higher than that of any regular observations currently available. Hence, the occurrence of the helical kink instability might be much more frequent than often thought. 

\item The analysis of the decay index excludes the torus instability as the eruption onset mechanism for all the flares studied, since the MFR is located well below the height where the decay index reaches its threshold value. However, for the eruptive flares we expect that the helical kink lifts the MFR into the torus-unstable height range. This will be studied in the future using MHD simulations.

\end{itemize}

To conclude, we found that an MFR existed prior to each investigated flare and that the twist number $\mathcal{T}_w$ is useful in 1) identifying the MFR in conjunction with the squashing factor $Q$, 2) forewarning an MFR eruption, and 3) locating the magnetic axis of a coherent MFR. 

\appendix

\section{Quality of NLFFF} \label{append:quality}
\begin{figure}
  \centering
  \includegraphics[width=\hsize]{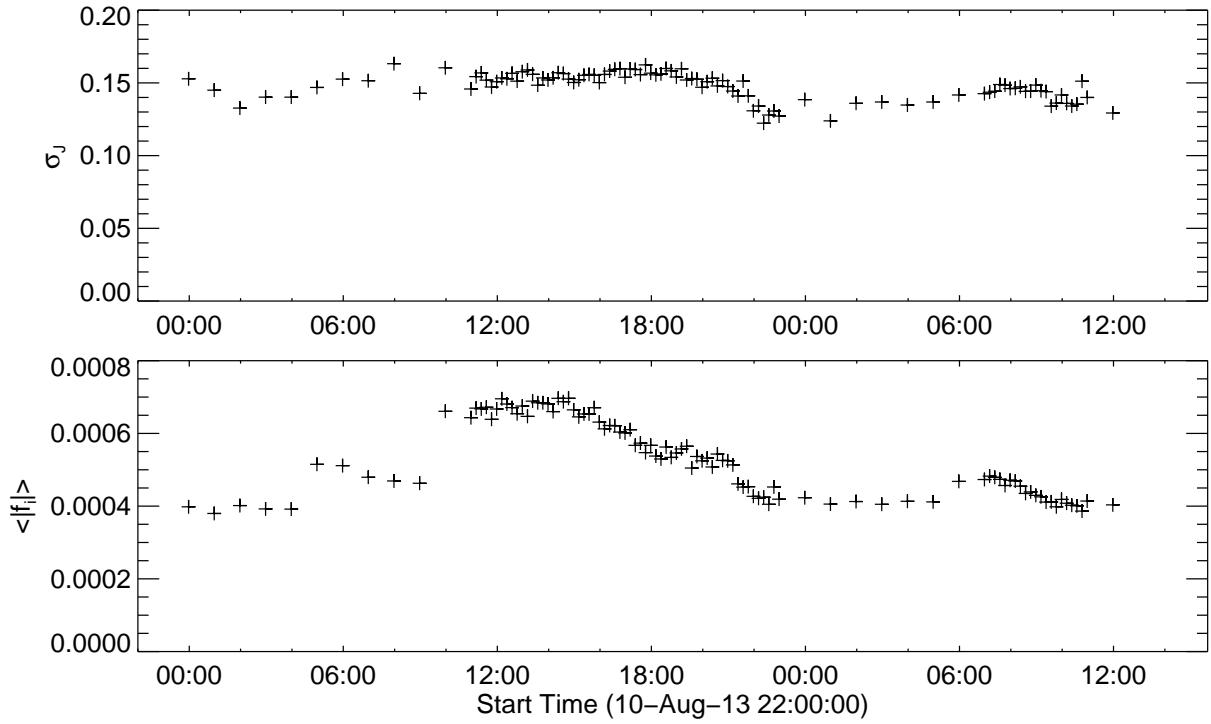} 
  \caption{\small Force-free and divergence-free conditions as gauged by $\sigma_J$ and $<|f_i|>$, respectively, for each reconstructed NLFFF used in the present study. \label{quality}}
\end{figure}

As the NLFFF extrapolation seeks an optimal solution that reduces $\mathbf{J}\times\mathbf{B}$ and $\nabla\cdot\mathbf{B}$ as much as possible under the constraint of the photospheric boundary, we calculate the following two dimensionless parameters to gauge the quality of the reconstructed NLFF fields \citep{Wheatland2000}
\begin{gather}
\sigma_J=\left(\sum_{i=1}^{n}\frac{|\mathbf{J}\times\mathbf{B}|_i}{B_i}\right)/\sum_{i=1}^{n}J_i, \\
<|f_i|>=\frac{1}{n}\sum_{i=1}^{n} \frac{|\nabla\cdot\mathbf{B}|_i\Delta V_i}{B_i\Delta S_i},
\end{gather} 
where the subscript $i$ runs over all the $n$ grid points within the computational box, and $\Delta V_i$ is the $i$th cell volume with the surface area $\Delta S_i$. Following \citet{Valori2013}, a cell volume of the cube of the spacing of the uniform grids is assigned to each internal grid points, but for those on the lateral surfaces, edges, and corners, the cell volume is reduced to 1/2, 1/4, and 1/8, respectively. $\sigma_J$ is the $J$-weighted average of $\sin\theta$, where $\theta$ is the angle between $\mathbf{J}$ and $\mathbf{B}$. $f_i$ measures the fractional change of flux in the volume $\Delta V_i$. For all the reconstructed NLFF fields used in the present study, $\sigma_J$ is typically 0.15 ($\theta\la10$~degrees on average), and $<|f_i|>$ is typically $5\times10^{-4}$ (Figure~\ref{quality}). Thus, both the force-free and divergence-free conditions are fairly satisfied.

\section{Error Analysis} \label{append:error}
Here we give a crude estimation of the uncertainties in helicity flux, Poynting flux and magnetic free energy, following a method to estimate the errors of helicity flux by \citet{liuy14}. For each helicity flux density map, we choose a small area ($32\times32$ pixels) in the quiet Sun region (rectangular region in Figure~\ref{bfield}(a)) and compute the median of the absolute values of the helicity flux density in this area. This median is taken as a `proxy' of the error of the helicity flux density, and the error of the helicity flux at that time is obtained by a multiplication of the square root of the number of pixels in the map, per the rule of uncertainty propagation for the summation of independent measurements. The same procedure is applied to Poynting flux. The resultant errors of helicity and Poynting flux are of order $10^{34}$ Mx$^2$~s$^{-1}$, $10^{24}$ erg~s$^{-1}$, respectively, less than 1 part in 100, which cannot be shown in Figure~\ref{helicity}. This result is consistent with \citet{liuy14}, who also performed a Monte-Carlo simulation that gives a similar result. 

For each NLFFF and corresponding potential field, we take the mean magnetic energy per unit column\footnote{For a data cube of $n_x \times n_y\times n_z$ pixels, a unit column has a length of $n_z$ and a cross section of $1 \times 1$.} in the same quiet-Sun area as the error for each unit column. The error in magnetic free energy per unit column is given by the root mean square of the NLFFF and potential field errors. The accumulated error in magnetic free energy is obtained by a multiplication of the number of pixels at the photospheric boundary, as each individual unit column is seldom independent of others. The error bars are shown in the bottom panel of Figure~\ref{helicity}. The typical relative error is 20\%, during the time when the AR produced the flares of interest, which is comparable to previous results with Monte-Carlo simulations \citep{Thalmann2008,jing09}. 

\section{Twist Number} \label{append:tw}

\subsection{$\mathcal{T}_g$, $\mathcal{T}_w$, and $N$} \label{append:Tg&Tw}

Let $\mathbf{x}(s)$ be a smooth, non-self-intersecting curve parametrized by arclength $s$ and $\mathbf{y}(s)$ a second such curve surrounding $\mathbf{x}$, jointly defining a ribbon. Points on $\mathbf{y}$ are expressed as $\mathbf{y}(s)=\mathbf{x}(s)+\varepsilon\rho(s)\uvec{V}(s)$, where $\uvec{V}(s)$ is a unit vector normal to $\mathbf{x}$, so that $\mathbf{y}$ is also parametrized by the arclength on $\mathbf{x}$, which we will refer to as the axis curve. With $\varphi$ denoting the rotation angle made by $\mathbf{y}$ about $\mathbf{x}$ and $\uvec{T}(s)$ denoting the unit tangent vector to $\mathbf{x}(s)$, the twist number of curve $\mathbf{y}$ about the axis $\mathbf{x}$ is given by \citep[Eq.~12 in][]{bp06}
\begin{equation}
\mathcal{T}_g = \frac{1}{2\pi}\int_\mathbf{x}\frac{d\varphi}{ds}ds 
 = \frac{1}{2\pi}\int_\mathbf{x}\uvec{T}(s)\cdot\uvec{V}(s)\times \frac{d\uvec{V}(s)}{ds}\,ds. \label{eq:Tg}
\end{equation}

Applying this equation to magnetic field lines, we have $\uvec{T} = \uvec{B} \equiv \mathbf{B}/B$. In order to derive the relationship between the general definition of twist number, $\mathcal{T}_g$, in a magnetic field and the alternative twist number $\mathcal{T}_w$ (Eq.~\ref{eq:Tw}), suggested in the literature to approximate $\mathcal{T}_g$ in the vicinity of $\mathbf{x}$ ($\varepsilon\ll1$), we express the rate of direction change $d\uvec{V}/ds$ through the structure (the derivatives) of the magnetic field. Denote the infinitesimally small distance between $\mathbf{x}$ and $\mathbf{y}$ at point $s$ by $\delta\mathbf{r}(s) = \varepsilon\rho(s)\uvec{V}(s) = \mathbf{y}(s)-\mathbf{x}(s)$. This quantity changes at a rate 
\begin{equation}
\frac{d}{ds}\delta\mathbf{r} = \frac{d\mathbf{y}}{ds}-\frac{d\mathbf{x}}{ds}.
\end{equation} 
We have $d\mathbf{x}/ds = \uvec{T}(s) = \uvec{B}(\mathbf{x}(s))$. Using arclength $s^\prime$ on $\mathbf{y}$ and denoting the unit tangent vector at $\mathbf{y}(s)$ by $\uvec{T}^\prime(s)$, we can write $d\mathbf{y}/ds = d\mathbf{y}/ds^\prime \cdot ds^\prime/ds = \uvec{T}^\prime(s) \cdot ds^\prime/ds = \uvec{B}(\mathbf{y}(s)) \cdot ds^\prime/ds$. For sufficiently small $\varepsilon$ the difference between the two arclengths can be neglected, $ds^\prime \approx ds$, so that 
\begin{equation}
\frac{d}{ds}\delta\mathbf{r} \simeq \uvec{B}(\mathbf{x}(s) + \delta\mathbf{r}(s)) -\uvec{B}(\mathbf{x}(s)).
\end{equation} 
Taylor expanding the first term yields
\begin{equation}
\frac{d}{ds}\delta\mathbf{r} \simeq \delta\mathbf{r} \cdot \frac{\partial\uvec{B}}{\partial\mathbf{r}}, 
\end{equation} 
so that 
\begin{equation}
\frac{d \uvec{V}}{ds} \simeq \uvec{V} \cdot \frac{\partial\uvec{B}}{\partial\mathbf{r}} - \frac{1}{\rho}\frac{d\rho}{ds}\uvec{V}, \label{eq:dvds}
\end{equation}
where all quantities are taken at position $\mathbf{x}(s)$. The local density of $\mathcal{T}_g$ can now be written as 
\begin{align}
\frac{d\mathcal{T}_g}{ds} &= \frac{1}{2\pi} \uvec{T} \cdot \uvec{V} \times \frac{d\uvec{V}}{ds} \nonumber \\
& \simeq \frac{1}{2\pi} \uvec{T}\cdot \uvec{V} \times \left(\uvec{V}\cdot\parb \right), \label{eq:dTgds}
\end{align}
In evaluating $\uvec{V}\cdot\parb$ we split $\partial\mathbf{B}/\partial\mathbf{r}$ into symmetric and antisymmetric parts (a similar technique is used, e.g., in the proof of the Cauchy-Helmholtz theorem for the form of the local velocity field in a fluid element; see, e.g., p.37, \citealt{Kiselev1999}):  
\begin{align*}
\left[ \uvec{V}\cdot\parb \right]_i &= \frac{1}{B}\widehat{V}_j\frac{\partial B_i}{\partial x_j} - \frac{B_i}{B^2}\widehat{V}_j\frac{\partial B}{\partial x_j} \\
& = \frac{1}{B}\widehat{V}_j \left[\frac{1}{2}\left(\frac{\partial B_i}{\partial x_j} + \frac{\partial B_j}{\partial x_i} \right) +  \frac{1}{2}\left(\frac{\partial B_i}{\partial x_j} - \frac{\partial B_j}{\partial x_i} \right)\right] - \frac{\widehat{T}_i}{B}\widehat{V}_j\frac{\partial B}{\partial x_j} \\
& = \frac{1}{B} \left(\mathcal{S}_{ij}\widehat{V}_j + \frac{1}{2}\mu_0\epsilon_{jik}\widehat{V}_j J_k\right) - \frac{\widehat{T}_i}{B}\widehat{V}_j\frac{\partial B}{\partial x_j}.
\end{align*}
Here $\mu_0\, \mathbf{J}=\nabla\times\mathbf{B}$ and $\mathcal{S}_{ij}\equiv [\mathbb{S}]_{ij}$ denotes the symmetric part of $\partial\mathbf{B}/\partial\mathbf{r}$.
Generally, the orthogonal triad of eigenvectors of $\mathbb{S}$ has an arbitrary orientation with respect to $\uvec{B}=\uvec{T}$, so that $\mathbb{S} \cdot \uvec{V} = c_1\, \uvec{T} + c_2\, \uvec{V} + c_3 \, \uvec{T} \times \uvec{V}$, where the coefficients $c_1$, $c_2$, and $c_3$ generally depend on both $\mathbb{S}$ and $\uvec{V}$.
Bearing also in mind that $\uvec{T} \cdot \uvec{V} \times \uvec{V} = \uvec{T} \cdot \uvec{V} \times \uvec{T} = 0 $ and $\uvec{T} \cdot \uvec{V} \times (\uvec{T} \times \uvec{V}) = 1 $, we conclude that
only the third term of $\mathbb{S} \cdot \uvec{V}$ and the antisymmetric part of $\partial\mathbf{B}/\partial\mathbf{r}$ contribute to ${d\mathcal{T}_g}/{ds}$ to give 
\begin{align}
\frac{d\mathcal{T}_g}{ds} & \simeq \frac{1}{2\pi B}\left[c_3 - \frac{1}{2}\mu_0\uvec{T}\cdot\uvec{V}\times(\uvec{V}\times\mathbf{J})\right] \nonumber\\
& = \frac{1}{2\pi B}\left[c_3+  \frac{1}{2}\mu_0 (\uvec{T}\cdot\mathbf{J}) \right]  \nonumber\\
& = \frac{c_3}{2\pi B} + \frac{\mu_0J_\parallel}{4\pi B}. \label{eq:dTgds_approx}
\end{align}
The contributions to $\mathcal{T}_g$ from the $c_3$ and $J_\parallel$ terms are referred to hereafter as $\mathcal{T}_{gc}$ and $\mathcal{T}_{gj}$, respectively, i.e.,
\begin{equation}
\mathcal{T}_{gc}=\int_\mathbf{x} \frac{c_3}{2\pi B}\,ds \quad\text{and}\quad  \mathcal{T}_{gj}=\int_\mathbf{x} \frac{\mu_0J_\parallel}{4\pi B}\,ds. \label{eq:Tgc&Tgj}
\end{equation}
Since all quantities in Eq.~\ref{eq:dTgds_approx} refer to the axis field line $\mathbf{x}(s)$, $\mathcal{T}_{gj}$ is \emph{not} equivalent to $\mathcal{T}_w$ (Eq.~\ref{eq:Tw}) which is evaluated at the field line of interest, $\mathbf{y}(s)$. Given a flux rope with a well defined axis, the term $\mathcal{T}_{gj}$ is identical for all field lines in the rope, while $\mathcal{T}_{gc}$ differs for different field lines because $\uvec{V}$ differs. In the limit $\varepsilon\to0$ the twist number $\mathcal{T}_w$ approaches $\mathcal{T}_{gj}$, so that 
\begin{equation}
\lim_{\varepsilon\to0}\mathcal{T}_w(\varepsilon) = \mathcal{T}_g - \int_\mathbf{x} \frac{c_3}{2\pi B}\,ds. \label{eq:Tg&Tw} 
\end{equation} 

It is now clear that $\mathcal{T}_w$ can serve as a reliable approximation of $\mathcal{T}_g$ only when two conditions are satisfied. First, the field line must be sufficiently close to the axis such that $J_\parallel/B$ on $\mathbf{x}$ and $\mathbf{y}$ are approximately equal \cite[as also required in][]{bp06}. Second, the contribution from $\mathbb{S}$ proportional to $c_3$ must be negligible. Particularly, in cylindrical symmetry, $B_r = 0$, $B_\phi=B_\phi(r)$, and $B_z=B_z(r)$, all elements of $\mathbb{S}$ vanish identically except
\begin{equation*}
S_{r\phi}\equiv\frac{1}{2}r\frac{\partial}{\partial r} \left(\frac{B_\phi}{r}\right). 
\end{equation*} 
For a smooth distribution of current density $J_\parallel(r)$ this term vanishes at the axis, so that $c_3=0$. For example, in a uniform-$\alpha$ force-free flux rope \citep{Lundquist1950}, $B_\phi = B_0J_1(\alpha r)$, hence $S_{r\phi} = -\alpha B_0 J_2(\alpha r)/2$, where $J_1$ and $J_2$ are Bessel functions of the first kind. Also, in a uniformly twisted flux rope \citep{Gold&Hoyle1960}, $B_\phi=br/(1+b^2r^2)$, hence $ S_{r\phi}= -b^3r^2/(1+b^2r^2)^2$. In both cases, $S_{r\phi}$ = 0 at the axis. Thus, one may use the ratio between the two terms in Eq.~\ref{eq:dTgds_approx}, $2c_3/\mu_0J_\parallel$, to evaluate locally how close a flux rope is to cylindrical symmetry. 

Of course, Eq.~\ref{eq:Tg} can be directly applied to a cylindrical flux tube. In this case, for all $s$, $\uvec{T}=\uvec{e}_z$, $\uvec{V} = \uvec{e}_r$, and 
\[\frac{d\uvec{V}}{dz} = \frac{d\phi}{dz}\,\uvec{e}_\phi.\] 
From the field line equation in cylindrical coordinates, $dr/B_r = rd\phi/B_\phi = dz/B_z = ds/B$, we have 
\begin{equation}
\mathcal{T}_g = \frac{1}{2\pi}\int d\phi = \frac{1}{2\pi}\int \frac{B_\phi(r)}{rB_z(r)}dz.
\end{equation}
By cylindrical symmetry, the radial distance of the field line is independent of $z$, so that the classical expression for the winding number $N$ about the $z$ axis in a cylinder of length $L_z$ is obtained,
\begin{equation}
\mathcal{T}_g(r) = \frac{1}{2\pi}\frac{L_zB_\phi(r)}{rB_z(r)} = N(r).\label{eq:N_Bphi}
\end{equation}

\subsection{Properties of $\mathcal{T}_w$}

The twist number $\mathcal{T}_w$ can straightforwardly be computed for any field line without resorting to the geometry of an MFR, providing a convenient means to characterize and quantify magnetic configurations through maps of $\mathcal{T}_w$ (``twist maps''). This must be done with some caution, however. The considerations of Appendix~\ref{append:Tg&Tw} show that $\mathcal{T}_w$ represents the classical meaning of twist---winding of field lines about an axis field line---only under certain conditions. An MFR structure possessing some degree of coherence must exist. In this case, $\mathcal{T}_w$ is a close approximation of its twist only in the vicinity of the magnetic axis and if the MFR is approximately cylindrically symmetric. The latter condition can be quantified from Eq.~\ref{eq:dTgds_approx} as $c_3\ll\mu_0J_\parallel/2$.

$\mathcal{T}_w$ is also useful in locating the magnetic axis of an MFR, the nontrivial requirement for the computation of $\mathcal{T}_g$. From Eq.~\ref{eq:Tw}, $\mathcal{T}_w$ may peak at any distance $r$ from the axis, depending on the radial profile of $J_\parallel/B$. For a force-free field this is $\alpha(r)$. If the MFR possesses some degree of cylindrical symmetry, $\mathcal{T}_w$ will have a local extremum at its axis, except in the special case that $J_\parallel/B$ is uniform in the vicinity of the axis. The uniform-$\alpha$ force-free rope \citep{Lundquist1950} has the current distribution $J_z(r)=\mu_0^{-1}\alpha B_0J_0(\alpha r)$, where $J_0$ is the zeroth-order Bessel function. Thus, $\mathcal{T}_w$ will peak at the axis of a force-free flux rope if $J_\parallel(r)$ is more peaked than this function and will have a minimum at the axis in the opposite case.

To demonstrate the usefulness of $\mathcal{T}_w$ in locating the magnetic axis of an MFR as well as its limitation in approximating $\mathcal{T}_g$, we consider again the cylindrically symmetric flux ropes, in which $\mathcal{T}_g=N$. For the Lundquist flux rope  $\mathcal{T}_w$ approaches $N$ at its axis, 
\begin{equation*}
\lim_{r\to 0}N(r)=\lim_{r\to 0} \frac{B_0J_1(\alpha r)}{2\pi rB_0J_0(\alpha r)}L_z= \frac{\alpha}{4\pi}L_z,
\end{equation*} 
but in general, 
\begin{eqnarray} 
\frac{N(r)}{\mathcal{T}_w(r)} &=& \frac{B_\phi L_z/2\pi rB_z}{\alpha|B|L_z/4\pi B_z} = \frac{2B_\phi}{\alpha r|B|} \nonumber \\
&=& \frac{2J_1(\alpha r)}{\alpha r\surd{J_0^2(\alpha r)+J_1^2(\alpha r)}}<1.
\end{eqnarray} 
For example, $\mathcal{T}_w$ overestimates $N$ by 25~percent at $r\approx2.5\alpha^{-1}$ in this rope.

In the uniformly twisted flux rope, we have $N = bL_z/2\pi$, where the constant $2\pi/b$ is the axial length of one field line turn, $L_z/N$, given by $b=d\phi/dz=B_\phi/rB_z=2\pi N/L_z$. With $B_\phi=br/(1+b^2r^2)$ and $B_z=1/(1+b^2r^2)$, we have 
\begin{align*}
\mathcal{T}_w &= \int_L\frac{(\nabla\times\mathbf{B})\cdot \mathbf{B}}{4\pi B^2}\,ds  =\int_{L_z}\frac{(\nabla\times\mathbf{B})\cdot \mathbf{B}}{4\pi BB_z}\,dz \\
&= \frac{1}{4\pi} \frac{2b}{\sqrt{1+b^2r^2}} L_z,
\end{align*} 
and hence
\begin{equation}
\frac{N}{\mathcal{T}_w}=\sqrt{1+b^2r^2}.
\end{equation} 
Again, $\mathcal{T}_w$ matches $N$ at the axis, but away from the axis $\mathcal{T}_w$ underestimates $N$, contrary to the case of the Lundquist rope. 

We have further checked the approach of $\mathcal{T}_w$ and $N$ at the flux rope axis and the location of the peak $\mathcal{T}_w$ for an approximately force-free Titov-D\'emoulin flux rope equilibrium, using two different choices for the toroidal current density $J_\mathrm{t}(r)$, one roughly uniform as in the original construction of the equilibrium \citep{titov&demoulin99}, the other with a distribution of $J_\mathrm{t}(r)$ strongly peaked at $r=0$. By construction, each small section of this toroidal flux rope possesses approximate cylindrical symmetry, especially for large aspect ratio. The computation was performed on a discrete grid, similar to the situation of a numerical extrapolation or an MHD simulation, in a vertical cut through the flux rope apex. For an aspect ratio of merely 3.4, $\mathcal{T}_w$ (Eq.~\ref{eq:Tw}) and $N$ (Eq.~\ref{eq:N_Bphi}) are found to agree to within 5~percent at the axis, where they also peak in the case of the peaked $J_\mathrm{t}(r)$. For the roughly uniform $J_\mathrm{t}(r)$, both peak at the periphery of the current channel in the center of the flux rope and are minimal at the magnetic axis, with similar radial profiles. 

Thus, assuming a coherent MFR, which possesses at least an approximate cylindrical symmetry, and excluding the special case of uniform $J_\parallel(r)/B(r)$, the magnetic axis is located at the local extremum of the $|\mathcal{T}_w|$ map, which either is a peak, or a dip enclosed by a ring of high $|\mathcal{T}_w|$ values. The flux ropes in the NLFFFs studied in this paper conform to the former case.

\subsection{Application to the Present MFR}
\begin{figure}
	\centering
	\includegraphics[width=\hsize]{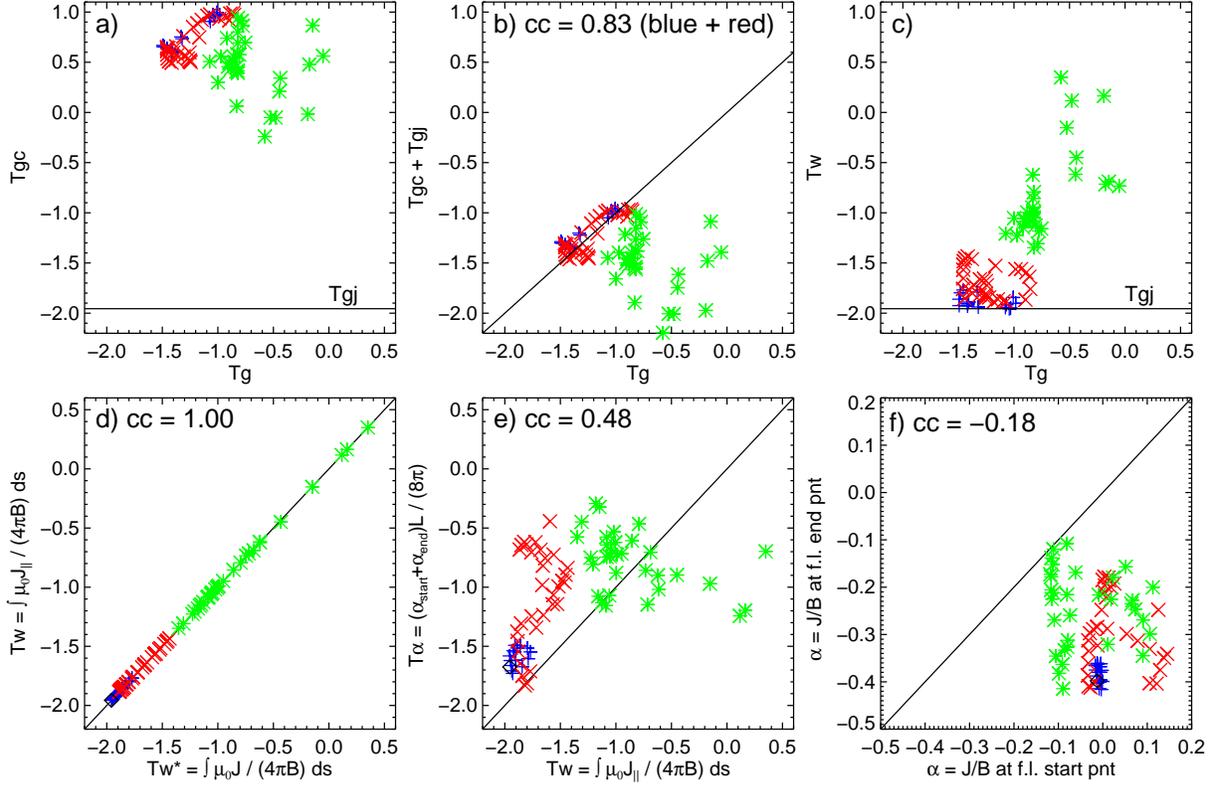}
	\caption{\small Twist number of field lines in Figure~\ref{rope3d} obtained with different approaches. The color code is the same as in Figure~\ref{rope3d}, except for the rainbow-colored field lines, whose twist number is shown in blue. $\mathcal{T}_g$ is obtained through Eq.~\ref{eq:Tg}. $\mathcal{T}_{gc}$ and $\mathcal{T}_{gj}$ are given by Eq.~\ref{eq:Tgc&Tgj}. $\mathcal{T}_w$ and $\mathcal{T}_w^*$ are calculated by integrating $J_\parallel/4\pi B$ (Eq.~\ref{eq:Tw}) and $J/4\pi B$ along the field line, respectively. $T_\alpha$ is calculated by multiplying the average $\alpha$ at the conjugate footpoints of the field line by the line length $L$.  \label{twvstn}}
\end{figure}

\begin{figure}
	\centering
	\includegraphics[width=\hsize]{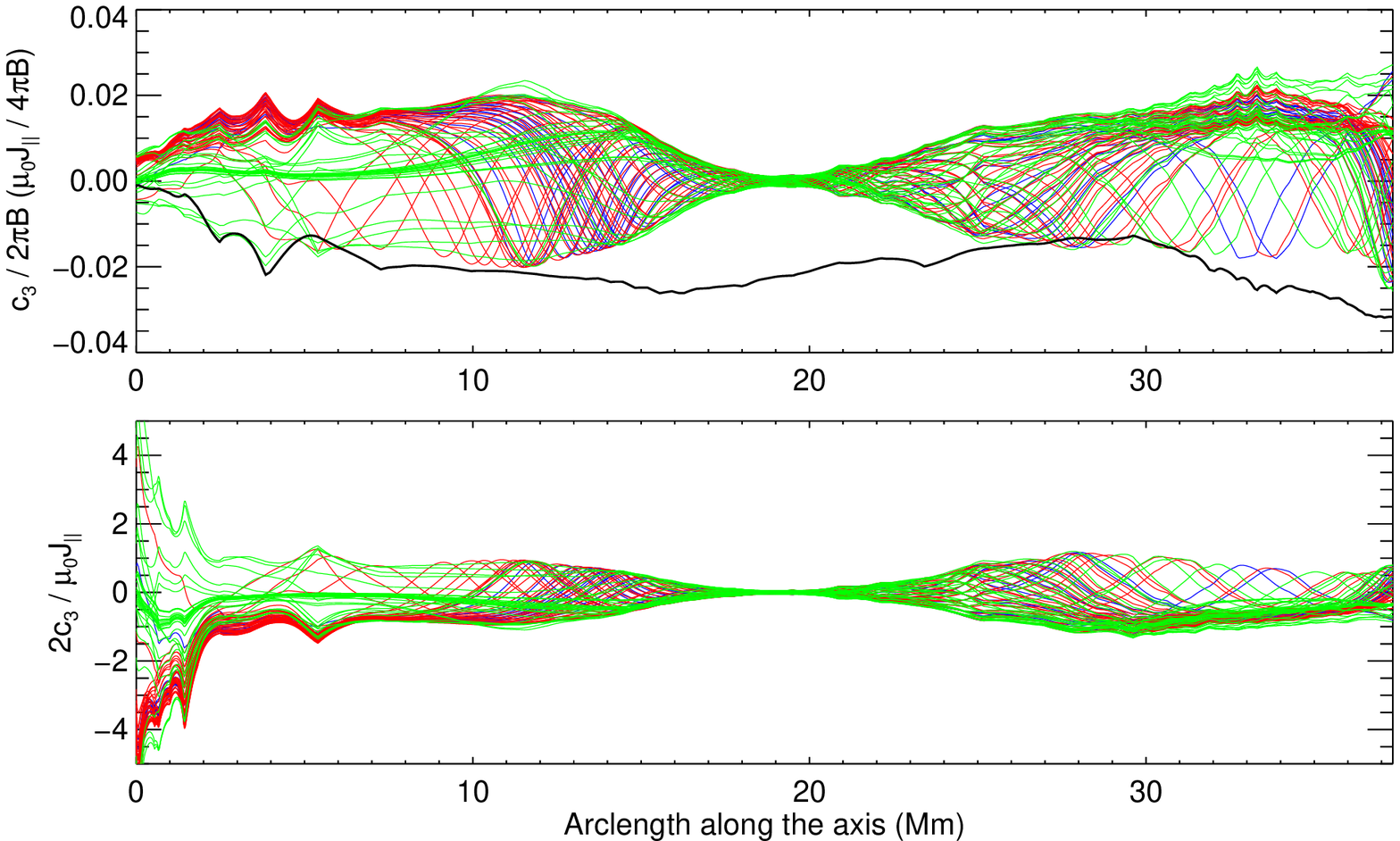}
	\caption{\small Local density of $\mathcal{T}_{gc}$ (color coded as in Figure~\ref{twvstn}) and $\mathcal{T}_{gj}$ (black; top panel) and their ratio (bottom panel) along the field lines in Figure~\ref{rope3d}. The abscissa is the arclength along the axis in units of Mm, starting from positive polarity, i.e., at the western end. \label{dtgds}}
\end{figure}

In the following we compare the different means of obtaining a twist number, using the flux rope shown in Figure~\ref{rope3d}. These are $\mathcal{T}_g$ from Eq.~\ref{eq:Tg}, its approximation broken into two terms, $\mathcal{T}_{gc}$ and $\mathcal{T}_{gj}$, from Eq.~\ref{eq:Tgc&Tgj}, $\mathcal{T}_w$ from Eq.~\ref{eq:Tw}, $\mathcal{T}_w^*$ from Eq.~\ref{eq:Tw_alpha}, and the simplified estimate of the latter, $\mathcal{T}_\alpha=(\alpha_\mathrm{start}+\alpha_\mathrm{end})L/8\pi$, obtained by multiplying the average value of $\alpha$ at the field line footpoints with the field line length $L$. 

Figure~\ref{twvstn}(a) and (b) examine the relationship between the exact $\mathcal{T}_g$ (Eq.~\ref{eq:Tg}) and the approximate one (integration of Eq.~\ref{eq:dTgds_approx} along the axis), as well as the individual terms in the latter (Eq.~\ref{eq:Tgc&Tgj}). The uniform value of $\mathcal{T}_{gj}$ is indicated by the horizontal line in (a). Both $\mathcal{T}_{gc}$ and $\mathcal{T}_{gj}$ contribute significantly to $\mathcal{T}_g$, however, none of them is individually correlated with $\mathcal{T}_g$, and near the axis they are of opposite sign. Their sum is highly correlated with $\mathcal{T}_g$ for all field lines of the inner two sets (blue and red), indicating that the approximation of small $\varepsilon$ underlying Eq.~\ref{eq:dTgds_approx} is here satisfied up to about one third of the rope's minor radius. As expected from the large radial distance and irregular appearance in the field line plot, there's no correlation for the remote (green) field lines.

Figure~\ref{twvstn}(c) shows that $\mathcal{T}_w$ and $\mathcal{T}_g$ are in only moderate agreement, due to two reasons. One first sees that the blue points agree rather well with $\mathcal{T}_{gj}$, whereas the red points here agree only moderately. This reflects the fact that the current density in our flux rope is considerably peaked at the axis (Figure~\ref{rope3d}(e)), so that $\mathcal{T}_w$, which is based purely on the local $J_\parallel(r)/B(r)$, begins to differ from $\mathcal{T}_{gj}$ at a smaller radial distance than the approximate $\mathcal{T}_g$ (Eq.~\ref{eq:dTgds_approx}) from the exact one (Eq.~\ref{eq:Tg}; Figure~\ref{twvstn}(b)). Second, even the blue points do not agree perfectly with $\mathcal{T}_g$, with ratios in the range $\mathcal{T}_w/\mathcal{T}_g \approx 1.3\mbox{--}2$. This implies that the term $\mathcal{T}_{gc}$ is not negligible for our flux rope, i.e., the rope as a whole is not close to cylindrical symmetry, consistent with its appearance in Figure~\ref{rope3d}. We will address this in detail below. 

The different forms of twist number not referring to an axis are compared in Figure~\ref{twvstn}(d) and (e). $\mathcal{T}_w$ and $\mathcal{T}_w^*$ are found to be highly correlated, indicating that the force-free condition is very well satisfied in the volume of the flux rope. On the other hand, the simplified variant, $\mathcal{T}_\alpha$, shows significant differences. Thus, deviations from force-freeness in the extrapolated field are particularly high at the footpoints of the field lines. This problem is also illustrated by the difference of $\alpha_\mathrm{start}$ and $\alpha_\mathrm{end}$, which can be substantial (Figure~\ref{twvstn}(f)), and has also been noted by \citet{Inoue2011} and \citet{Chintzoglou2015}. Further, we found that the correlation between $T_\alpha$ and $T_g$ becomes much worse when $\alpha$ at the footpoints is obtained by $J_z/B_z$ rather than by $J/B$. This may be due to the large uncertainties associated with the transverse field components $B_x$ and $B_y$ in general and those associated with the weak $B_z$ near the PIL in particular. 

The local density of $\mathcal{T}_{gj}$ and $\mathcal{T}_{gc}$ is further examined in Figure~\ref{dtgds}. $c_3/2\pi B$ is minimal and much smaller than $\mu_0J_\parallel/4\pi B$ in the middle section of the flux rope, which thus is nearly cylindrically symmetric. The fact that $c_3\approx0$ in the middle section indicates clearly that the magnetic axis of the forming flux rope was correctly located by the maximum of $|\mathcal{T}_w|$ in our twist maps.

\acknowledgments We thank the anonymous referee for constructive suggestions, the SDO team for the vector magnetic field and EUV imaging data, BBSO staff for the NST observations, P.W. Schuck for the flow tracking code, and Y. Guo and T.~T{\"o}r{\"o}k for helpful comments. RL acknowledges the Thousand Young Talents Program of China, NSFC 41222031 and 41474151, and NSF AGS-1153226. This work was also supported by NSFC 41131065 and 41121003, 973 key project 2011CB811403, CAS Key Research Program KZZD-EW-01-4, the fundamental research funds for the central universities WK2080000031. BK acknowledges support by the CAS under grant no.~2012T1J0017 and by the DFG. VT was supported by the NSF SHINE program. CL, HW and YX are supported by NSF under grants AGS 1348513 and 1408703, and by NASA under grants NNX13AG13G and NNX13AF76G


\begin{thebibliography}{}
\expandafter\ifx\csname natexlab\endcsname\relax\def\natexlab#1{#1}\fi

\bibitem[{{Alexander} {et~al.}(2006){Alexander}, {Liu}, \&
  {Gilbert}}]{alexander06}
{Alexander}, D., {Liu}, R., \& {Gilbert}, H.~R. 2006, \apj, 653, 719

\bibitem[{{Antiochos} {et~al.}(1999){Antiochos}, {DeVore}, \&
  {Klimchuk}}]{adk99}
{Antiochos}, S.~K., {DeVore}, C.~R., \& {Klimchuk}, J.~A. 1999, \apj, 510, 485

\bibitem[{{Baty}(2001)}]{Baty2001}
{Baty}, H. 2001, \aap, 367, 321

\bibitem[{{Baty} \& {Heyvaerts}(1996)}]{baty&heyvaerts96}
{Baty}, H., \& {Heyvaerts}, J. 1996, \aap, 308, 935

\bibitem[{{Berger}(1984)}]{berger84}
{Berger}, M.~A. 1984, Geophysical and Astrophysical Fluid Dynamics, 30, 79

\bibitem[{{Berger} \& {Prior}(2006)}]{bp06}
{Berger}, M.~A., \& {Prior}, C. 2006, Journal of Physics A Mathematical
  General, 39, 8321

\bibitem[{{Bobra} {et~al.}(2014){Bobra}, {Sun}, {Hoeksema}, {Turmon}, {Liu},
  {Hayashi}, {Barnes}, \& {Leka}}]{Bobra2014}
{Bobra}, M.~G., {Sun}, X., {Hoeksema}, J.~T., {et~al.} 2014, \solphys, 289,
  3549

\bibitem[{{Bobra} {et~al.}(2008){Bobra}, {van Ballegooijen}, \&
  {DeLuca}}]{bobra08}
{Bobra}, M.~G., {van Ballegooijen}, A.~A., \& {DeLuca}, E.~E. 2008, \apj, 672,
  1209

\bibitem[{{Canou} {et~al.}(2009){Canou}, {Amari}, {Bommier}, {Schmieder},
  {Aulanier}, \& {Li}}]{canou09}
{Canou}, A., {Amari}, T., {Bommier}, V., {et~al.} 2009, \apjl, 693, L27

\bibitem[{{Cheng} {et~al.}(2014){Cheng}, {Ding}, {Zhang}, {Sun}, {Guo}, {Wang},
  {Kliem}, \& {Deng}}]{Cheng2014}
{Cheng}, X., {Ding}, M.~D., {Zhang}, J., {et~al.} 2014, \apj, 789, 93

\bibitem[{{Chintzoglou} {et~al.}(2015){Chintzoglou}, {Patsourakos}, \&
  {Vourlidas}}]{Chintzoglou2015}
{Chintzoglou}, G., {Patsourakos}, S., \& {Vourlidas}, A. 2015, ArXiv e-prints,
  arXiv:1507.01165

\bibitem[{{Cho} {et~al.}(2009){Cho}, {Lee}, {Bong}, {Kim}, {Joshi}, \&
  {Park}}]{cho09}
{Cho}, K.-S., {Lee}, J., {Bong}, S.-C., {et~al.} 2009, \apj, 703, 1

\bibitem[{{Einaudi} \& {van Hoven}(1983)}]{einaudi&vanhoven83}
{Einaudi}, G., \& {van Hoven}, G. 1983, \solphys, 88, 163

\bibitem[{{Fan}(2005)}]{fan05}
{Fan}, Y. 2005, \apj, 630, 543

\bibitem[{{Forbes} \& {Priest}(1995)}]{forbes&priest95}
{Forbes}, T.~G., \& {Priest}, E.~R. 1995, \apj, 446, 377

\bibitem[{{Gary} \& {Moore}(2004)}]{gary&moore04}
{Gary}, G.~A., \& {Moore}, R.~L. 2004, \apj, 611, 545

\bibitem[{{Gilchrist} {et~al.}(2012){Gilchrist}, {Wheatland}, \&
  {Leka}}]{Gilchrist12}
{Gilchrist}, S.~A., {Wheatland}, M.~S., \& {Leka}, K.~D. 2012, \solphys, 276,
  133

\bibitem[{{Gold} \& {Hoyle}(1960)}]{Gold&Hoyle1960}
{Gold}, T., \& {Hoyle}, F. 1960, \mnras, 120, 89

\bibitem[{{Green} {et~al.}(2007){Green}, {Kliem}, {T{\"o}r{\"o}k}, {van
  Driel-Gesztelyi}, \& {Attrill}}]{green07}
{Green}, L.~M., {Kliem}, B., {T{\"o}r{\"o}k}, T., {van Driel-Gesztelyi}, L., \&
  {Attrill}, G.~D.~R. 2007, \solphys, 246, 365

\bibitem[{{Guo} {et~al.}(2013){Guo}, {Ding}, {Cheng}, {Zhao}, \&
  {Pariat}}]{YGuo&al2013}
{Guo}, Y., {Ding}, M.~D., {Cheng}, X., {Zhao}, J., \& {Pariat}, E. 2013, \apj,
  779, 157

\bibitem[{{Guo} {et~al.}(2010{\natexlab{a}}){Guo}, {Ding}, {Schmieder}, {Li},
  {T{\"o}r{\"o}k}, \& {Wiegelmann}}]{guo10b}
{Guo}, Y., {Ding}, M.~D., {Schmieder}, B., {et~al.} 2010{\natexlab{a}}, \apjl,
  725, L38

\bibitem[{{Guo} {et~al.}(2010{\natexlab{b}}){Guo}, {Schmieder}, {D{\'e}moulin},
  {Wiegelmann}, {Aulanier}, {T{\"o}r{\"o}k}, \& {Bommier}}]{guo10a}
{Guo}, Y., {Schmieder}, B., {D{\'e}moulin}, P., {et~al.} 2010{\natexlab{b}},
  \apj, 714, 343

\bibitem[{{Hannah} \& {Kontar}(2012)}]{hk12}
{Hannah}, I.~G., \& {Kontar}, E.~P. 2012, \aap, 539, A146

\bibitem[{{Hoeksema} {et~al.}(2014){Hoeksema}, {Liu}, {Hayashi}, {Sun},
  {Schou}, {Couvidat}, {Norton}, {Bobra}, {Centeno}, {Leka}, {Barnes}, \&
  {Turmon}}]{hoeksema14}
{Hoeksema}, J.~T., {Liu}, Y., {Hayashi}, K., {et~al.} 2014, \solphys, 289, 3483

\bibitem[{{Hood} \& {Priest}(1979)}]{hood&priest79}
{Hood}, A.~W., \& {Priest}, E.~R. 1979, \solphys, 64, 303

\bibitem[{{Hood} \& {Priest}(1981)}]{hood&priest81}
---. 1981, Geophysical and Astrophysical Fluid Dynamics, 17, 297

\bibitem[{{Inoue} {et~al.}(2011){Inoue}, {Kusano}, {Magara}, {Shiota}, \&
  {Yamamoto}}]{Inoue2011}
{Inoue}, S., {Kusano}, K., {Magara}, T., {Shiota}, D., \& {Yamamoto}, T.~T.
  2011, \apj, 738, 161

\bibitem[{{Isenberg} \& {Forbes}(2007)}]{isenberg&forbes07}
{Isenberg}, P.~A., \& {Forbes}, T.~G. 2007, \apj, 670, 1453

\bibitem[{{Ji} {et~al.}(2003){Ji}, {Wang}, {Schmahl}, {Moon}, \&
  {Jiang}}]{ji03}
{Ji}, H., {Wang}, H., {Schmahl}, E.~J., {Moon}, Y.-J., \& {Jiang}, Y. 2003,
  \apjl, 595, L135

\bibitem[{{Jing} {et~al.}(2009){Jing}, {Chen}, {Wiegelmann}, {Xu}, {Park}, \&
  {Wang}}]{jing09}
{Jing}, J., {Chen}, P.~F., {Wiegelmann}, T., {et~al.} 2009, \apj, 696, 84

\bibitem[{{Jing} {et~al.}(2010){Jing}, {Yuan}, {Wiegelmann}, {Xu}, {Liu}, \&
  {Wang}}]{jing10}
{Jing}, J., {Yuan}, Y., {Wiegelmann}, T., {et~al.} 2010, \apjl, 719, L56

\bibitem[{{Karlick{\'y}} \& {Kliem}(2010)}]{karlicky&kliem10}
{Karlick{\'y}}, M., \& {Kliem}, B. 2010, \solphys, 266, 71

\bibitem[{{Kiselev} {et~al.}(1999){Kiselev}, {Vorozhtsov}, \&
  {Fomin}}]{Kiselev1999}
{Kiselev}, S.~P., {Vorozhtsov}, E., \& {Fomin}, V.~M. 1999, Foundations of
  Fluid Mechanics with Applications: Problem Solving Using
  Mathematica{\textregistered}, Modeling and Simulation in Science, Engineering
  and Technology (Birkh\"{a}user Boston)

\bibitem[{{Kliem} {et~al.}(2010){Kliem}, {Linton}, {T{\"o}r{\"o}k}, \&
  {Karlick{\'y}}}]{kliem10}
{Kliem}, B., {Linton}, M.~G., {T{\"o}r{\"o}k}, T., \& {Karlick{\'y}}, M. 2010,
  \solphys, 266, 91

\bibitem[{{Kliem} {et~al.}(2013){Kliem}, {Su}, {van Ballegooijen}, \&
  {DeLuca}}]{Kliem&al2013}
{Kliem}, B., {Su}, Y.~N., {van Ballegooijen}, A.~A., \& {DeLuca}, E.~E. 2013,
  \apj, 779, 129

\bibitem[{{Kliem} \& {T{\"o}r{\"o}k}(2006)}]{kliem&torok06}
{Kliem}, B., \& {T{\"o}r{\"o}k}, T. 2006, \prl, 96, 255002

\bibitem[{{Kliem} {et~al.}(2012){Kliem}, {T{\"o}r{\"o}k}, \&
  {Thompson}}]{Kliem2012}
{Kliem}, B., {T{\"o}r{\"o}k}, T., \& {Thompson}, W.~T. 2012, \solphys, 281, 137

\bibitem[{{Kliem} {et~al.}(2014){Kliem}, {T{\"o}r{\"o}k}, {Titov}, {Lionello},
  {Linker}, {Liu}, {Liu}, \& {Wang}}]{Kliem&al2014}
{Kliem}, B., {T{\"o}r{\"o}k}, T., {Titov}, V.~S., {et~al.} 2014, \apj, 792, 107

\bibitem[{{Kumar} \& {Cho}(2014)}]{kumar&cho14}
{Kumar}, P., \& {Cho}, K.-S. 2014, \aap, 572, A83

\bibitem[{{Kumar} {et~al.}(2012){Kumar}, {Cho}, {Bong}, {Park}, \&
  {Kim}}]{kumar12}
{Kumar}, P., {Cho}, K.-S., {Bong}, S.-C., {Park}, S.-H., \& {Kim}, Y.~H. 2012,
  \apj, 746, 67

\bibitem[{{Kusano} {et~al.}(2002){Kusano}, {Maeshiro}, {Yokoyama}, \&
  {Sakurai}}]{kusano02}
{Kusano}, K., {Maeshiro}, T., {Yokoyama}, T., \& {Sakurai}, T. 2002, \apj, 577,
  501

\bibitem[{{LaBonte} {et~al.}(2007){LaBonte}, {Georgoulis}, \&
  {Rust}}]{labonte07}
{LaBonte}, B.~J., {Georgoulis}, M.~K., \& {Rust}, D.~M. 2007, \apj, 671, 955

\bibitem[{{Leamon} {et~al.}(2003){Leamon}, {Canfield}, {Blehm}, \&
  {Pevtsov}}]{leamon03}
{Leamon}, R.~J., {Canfield}, R.~C., {Blehm}, Z., \& {Pevtsov}, A.~A. 2003,
  \apjl, 596, L255

\bibitem[{{Leka} \& {Barnes}(2007)}]{leka&barnes07}
{Leka}, K.~D., \& {Barnes}, G. 2007, \apj, 656, 1173

\bibitem[{{Leka} {et~al.}(2005){Leka}, {Fan}, \& {Barnes}}]{leka05}
{Leka}, K.~D., {Fan}, Y., \& {Barnes}, G. 2005, \apj, 626, 1091

\bibitem[{{Lemen} {et~al.}(2012){Lemen}, {Title}, {Akin}, {et~al.}}]{lemen12}
{Lemen}, J.~R., {Title}, A.~M., {Akin}, D.~J., {et~al.} 2012, \solphys, 275, 17

\bibitem[{{Lin} {et~al.}(2003){Lin}, {Soon}, \& {Baliunas}}]{lin03}
{Lin}, J., {Soon}, W., \& {Baliunas}, S.~L. 2003, \nar, 47, 53

\bibitem[{{Liu} \& {Alexander}(2009)}]{liu&alexander09}
{Liu}, R., \& {Alexander}, D. 2009, \apj, 697, 999

\bibitem[{{Liu} {et~al.}(2007){Liu}, {Alexander}, \& {Gilbert}}]{liu07}
{Liu}, R., {Alexander}, D., \& {Gilbert}, H.~R. 2007, \apj, 661, 1260

\bibitem[{{Liu} {et~al.}(2012){Liu}, {Kliem}, {T{\"o}r{\"o}k}, {Liu}, {Titov},
  {Lionello}, {Linker}, \& {Wang}}]{liu12}
{Liu}, R., {Kliem}, B., {T{\"o}r{\"o}k}, T., {et~al.} 2012, \apj, 756, 59

\bibitem[{{Liu} {et~al.}(2010){Liu}, {Liu}, {Wang}, {Deng}, \& {Wang}}]{liu10}
{Liu}, R., {Liu}, C., {Wang}, S., {Deng}, N., \& {Wang}, H. 2010, \apjl, 725,
  L84

\bibitem[{{Liu} {et~al.}(2014{\natexlab{a}}){Liu}, {Titov}, {Gou}, {Wang},
  {Liu}, \& {Wang}}]{liu14}
{Liu}, R., {Titov}, V.~S., {Gou}, T., {et~al.} 2014{\natexlab{a}}, \apj, 790, 8

\bibitem[{{Liu} {et~al.}(2014{\natexlab{b}}){Liu}, {Hoeksema}, {Bobra},
  {Hayashi}, {Schuck}, \& {Sun}}]{liuy14}
{Liu}, Y., {Hoeksema}, J.~T., {Bobra}, M., {et~al.} 2014{\natexlab{b}}, \apj,
  785, 13

\bibitem[{{Liu} \& {Schuck}(2012)}]{liuy12}
{Liu}, Y., \& {Schuck}, P.~W. 2012, \apj, 761, 105

\bibitem[{Lundquist(1950)}]{Lundquist1950}
Lundquist, S. 1950, Ark. Fys, 2, 361

\bibitem[{{Maeshiro} {et~al.}(2005){Maeshiro}, {Kusano}, {Yokoyama}, \&
  {Sakurai}}]{maeshiro05}
{Maeshiro}, T., {Kusano}, K., {Yokoyama}, T., \& {Sakurai}, T. 2005, \apj, 620,
  1069

\bibitem[{{Magara}(2006)}]{magara06}
{Magara}, T. 2006, \apj, 653, 1499

\bibitem[{{Moore} {et~al.}(2001){Moore}, {Sterling}, {Hudson}, \&
  {Lemen}}]{moore01}
{Moore}, R.~L., {Sterling}, A.~C., {Hudson}, H.~S., \& {Lemen}, J.~R. 2001,
  \apj, 552, 833

\bibitem[{{Pariat} \& {D{\'e}moulin}(2012)}]{Pariat&Demoulin12}
{Pariat}, E., \& {D{\'e}moulin}, P. 2012, \aap, 541, A78

\bibitem[{{Park} {et~al.}(2010){Park}, {Chae}, {Jing}, {Tan}, \&
  {Wang}}]{park10}
{Park}, S.-H., {Chae}, J., {Jing}, J., {Tan}, C., \& {Wang}, H. 2010, \apj,
  720, 1102

\bibitem[{{Park} {et~al.}(2008){Park}, {Lee}, {Choe}, {Chae}, {Jeong}, {Yang},
  {Jing}, \& {Wang}}]{park08}
{Park}, S.-H., {Lee}, J., {Choe}, G.~S., {et~al.} 2008, \apj, 686, 1397

\bibitem[{{Pesnell} {et~al.}(2012){Pesnell}, {Thompson}, \&
  {Chamberlin}}]{pesnell12}
{Pesnell}, W.~D., {Thompson}, B.~J., \& {Chamberlin}, P.~C. 2012, \solphys,
  275, 3

\bibitem[{{Phillips} {et~al.}(2005){Phillips}, {MacNeice}, \&
  {Antiochos}}]{phillips05}
{Phillips}, A.~D., {MacNeice}, P.~J., \& {Antiochos}, S.~K. 2005, \apjl, 624,
  L129

\bibitem[{{Qiu} {et~al.}(2007){Qiu}, {Hu}, {Howard}, \& {Yurchyshyn}}]{qiu07}
{Qiu}, J., {Hu}, Q., {Howard}, T.~A., \& {Yurchyshyn}, V.~B. 2007, \apj, 659,
  758

\bibitem[{{R{\'e}gnier} \& {Amari}(2004)}]{ra04}
{R{\'e}gnier}, S., \& {Amari}, T. 2004, \aap, 425, 345

\bibitem[{{Romano} {et~al.}(2003){Romano}, {Contarino}, \&
  {Zuccarello}}]{romano03}
{Romano}, P., {Contarino}, L., \& {Zuccarello}, F. 2003, \solphys, 214, 313

\bibitem[{{Rust} \& {LaBonte}(2005)}]{rust&labonte05}
{Rust}, D.~M., \& {LaBonte}, B.~J. 2005, \apj, 622, L69

\bibitem[{{Savcheva} {et~al.}(2012){Savcheva}, {Pariat}, {van Ballegooijen},
  {Aulanier}, \& {DeLuca}}]{Savcheva&al2012}
{Savcheva}, A., {Pariat}, E., {van Ballegooijen}, A., {Aulanier}, G., \&
  {DeLuca}, E. 2012, \apj, 750, 15

\bibitem[{{Schindler}(2006)}]{Schindler2006}
{Schindler}, K. 2006, {Physics of Space Plasma Activity} (Cambridge, UK:
  Cambridge University Press, 522 p.), doi:10.2277/0521858976

\bibitem[{{Schuck}(2008)}]{schuck08}
{Schuck}, P.~W. 2008, \apj, 683, 1134

\bibitem[{{Srivastava} {et~al.}(2010){Srivastava}, {Zaqarashvili}, {Kumar}, \&
  {Khodachenko}}]{srivastava10}
{Srivastava}, A.~K., {Zaqarashvili}, T.~V., {Kumar}, P., \& {Khodachenko},
  M.~L. 2010, \apj, 715, 292

\bibitem[{{Su} {et~al.}(2011){Su}, {Surges}, {van Ballegooijen}, {DeLuca}, \&
  {Golub}}]{su11}
{Su}, Y., {Surges}, V., {van Ballegooijen}, A., {DeLuca}, E., \& {Golub}, L.
  2011, \apj, 734, 53

\bibitem[{{Sun} {et~al.}(2012){Sun}, {Hoeksema}, {Liu}, {Wiegelmann},
  {Hayashi}, {Chen}, \& {Thalmann}}]{sun12}
{Sun}, X., {Hoeksema}, J.~T., {Liu}, Y., {et~al.} 2012, \apj, 748, 77

\bibitem[{{Thalmann} \& {Wiegelmann}(2008)}]{tw08}
{Thalmann}, J.~K., \& {Wiegelmann}, T. 2008, \aap, 484, 495

\bibitem[{{Thalmann} {et~al.}(2008){Thalmann}, {Wiegelmann}, \&
  {Raouafi}}]{Thalmann2008}
{Thalmann}, J.~K., {Wiegelmann}, T., \& {Raouafi}, N.-E. 2008, \aap, 488, L71

\bibitem[{{Titov}(2007)}]{titov07}
{Titov}, V.~S. 2007, \apj, 660, 863

\bibitem[{{Titov} \& {D{\'e}moulin}(1999)}]{titov&demoulin99}
{Titov}, V.~S., \& {D{\'e}moulin}, P. 1999, \aap, 351, 707

\bibitem[{{Titov} {et~al.}(2002){Titov}, {Hornig}, \& {D{\'e}moulin}}]{titov02}
{Titov}, V.~S., {Hornig}, G., \& {D{\'e}moulin}, P. 2002, Journal of
  Geophysical Research (Space Physics), 107, 1164

\bibitem[{{T{\"o}r{\"o}k} \& {Kliem}(2003)}]{torok&kliem03}
{T{\"o}r{\"o}k}, T., \& {Kliem}, B. 2003, \aap, 406, 1043

\bibitem[{{T{\"o}r{\"o}k} \& {Kliem}(2005)}]{torok&kliem05}
---. 2005, \apj, 630, L97

\bibitem[{T{\"o}r{\"o}k \& Kliem(2007)}]{Torok2007a}
T{\"o}r{\"o}k, T., \& Kliem, B. 2007, Astronomische Nachrichten, 328, 743

\bibitem[{{T{\"o}r{\"o}k} {et~al.}(2004){T{\"o}r{\"o}k}, {Kliem}, \&
  {Titov}}]{torok04}
{T{\"o}r{\"o}k}, T., {Kliem}, B., \& {Titov}, V.~S. 2004, \aap, 413, L27

\bibitem[{{Tziotziou} {et~al.}(2013){Tziotziou}, {Georgoulis}, \&
  {Liu}}]{Tziotziou13}
{Tziotziou}, K., {Georgoulis}, M.~K., \& {Liu}, Y. 2013, \apj, 772, 115

\bibitem[{{Tziotziou} {et~al.}(2012){Tziotziou}, {Georgoulis}, \&
  {Raouafi}}]{Tziotziou12}
{Tziotziou}, K., {Georgoulis}, M.~K., \& {Raouafi}, N.-E. 2012, \apjl, 759, L4

\bibitem[{{Valori} {et~al.}(2013){Valori}, {D{\'e}moulin}, {Pariat}, \&
  {Masson}}]{Valori2013}
{Valori}, G., {D{\'e}moulin}, P., {Pariat}, E., \& {Masson}, S. 2013, \aap,
  553, A38

\bibitem[{{Valori} {et~al.}(2005){Valori}, {Kliem}, \& {Keppens}}]{valori05}
{Valori}, G., {Kliem}, B., \& {Keppens}, R. 2005, \aap, 433, 335

\bibitem[{{Valori} {et~al.}(2010){Valori}, {Kliem}, {T{\"o}r{\"o}k}, \&
  {Titov}}]{valori10}
{Valori}, G., {Kliem}, B., {T{\"o}r{\"o}k}, T., \& {Titov}, V.~S. 2010, \aap,
  519, A44

\bibitem[{{van Tend} \& {Kuperus}(1978)}]{vanTend&Kuperus1978}
{van Tend}, W., \& {Kuperus}, M. 1978, \solphys, 59, 115

\bibitem[{{Vrsnak} {et~al.}(1991){Vrsnak}, {Ruzdjak}, \& {Rompolt}}]{vrsnak91}
{Vrsnak}, B., {Ruzdjak}, V., \& {Rompolt}, B. 1991, \solphys, 136, 151

\bibitem[{{Vrsnak} {et~al.}(1993){Vrsnak}, {Ruzdjak}, {Rompolt}, {Rosa}, \&
  {Zlobec}}]{vrsnak93}
{Vrsnak}, B., {Ruzdjak}, V., {Rompolt}, B., {Rosa}, D., \& {Zlobec}, P. 1993,
  \solphys, 146, 147

\bibitem[{{Wang} {et~al.}(2015){Wang}, {Cao}, {Liu}, {Xu}, {Liu}, {Zeng},
  {Chae}, \& {Ji}}]{wang15}
{Wang}, H., {Cao}, W., {Liu}, C., {et~al.} 2015, Nature Communications, 6, 7008

\bibitem[{{Wheatland} {et~al.}(2000){Wheatland}, {Sturrock}, \&
  {Roumeliotis}}]{Wheatland2000}
{Wheatland}, M.~S., {Sturrock}, P.~A., \& {Roumeliotis}, G. 2000, \apj, 540,
  1150

\bibitem[{{Wiegelmann}(2004)}]{wiegelmann04}
{Wiegelmann}, T. 2004, \solphys, 219, 87

\bibitem[{{Wiegelmann} {et~al.}(2006{\natexlab{a}}){Wiegelmann}, {Inhester},
  {Kliem}, {Valori}, \& {Neukirch}}]{wiegelmann06aa}
{Wiegelmann}, T., {Inhester}, B., {Kliem}, B., {Valori}, G., \& {Neukirch}, T.
  2006{\natexlab{a}}, \aap, 453, 737

\bibitem[{{Wiegelmann} {et~al.}(2006{\natexlab{b}}){Wiegelmann}, {Inhester}, \&
  {Sakurai}}]{wiegelmann06}
{Wiegelmann}, T., {Inhester}, B., \& {Sakurai}, T. 2006{\natexlab{b}},
  \solphys, 233, 215

\bibitem[{{Wiegelmann} {et~al.}(2012){Wiegelmann}, {Thalmann}, {Inhester},
  {Tadesse}, {Sun}, \& {Hoeksema}}]{wiegelmann12}
{Wiegelmann}, T., {Thalmann}, J.~K., {Inhester}, B., {et~al.} 2012, \solphys,
  281, 37

\bibitem[{{Williams} {et~al.}(2005){Williams}, {T{\"o}r{\"o}k}, {D{\'e}moulin},
  {van Driel-Gesztelyi}, \& {Kliem}}]{williams05}
{Williams}, D.~R., {T{\"o}r{\"o}k}, T., {D{\'e}moulin}, P., {van
  Driel-Gesztelyi}, L., \& {Kliem}, B. 2005, \apjl, 628, L163

\bibitem[{{Yang} {et~al.}(2012){Yang}, {Jiang}, {Bi}, {Li}, {Hong}, {Yang},
  {Zheng}, \& {Yang}}]{yang12}
{Yang}, J., {Jiang}, Y., {Bi}, Y., {et~al.} 2012, \apj, 749, 12

\bibitem[{{Zhang} {et~al.}(2012){Zhang}, {Cheng}, \& {Ding}}]{zhang12}
{Zhang}, J., {Cheng}, X., \& {Ding}, M.-D. 2012, Nature Communications, 3, id.
  747

\bibitem[{{Zhang} {et~al.}(2001){Zhang}, {Dere}, {Howard}, {Kundu}, \&
  {White}}]{JZhang&al2001}
{Zhang}, J., {Dere}, K.~P., {Howard}, R.~A., {Kundu}, M.~R., \& {White}, S.~M.
  2001, \apj, 559, 452

\bibitem[{{Zhang} \& {Flyer}(2008)}]{Zhang&Flyer08}
{Zhang}, M., \& {Flyer}, N. 2008, \apj, 683, 1160

\bibitem[{{Zhu} \& {Alexander}(2014)}]{Zhu&Alexander2014}
{Zhu}, C., \& {Alexander}, D. 2014, \solphys, 289, 279

\bibitem[{{Zuccarello} {et~al.}(2009){Zuccarello}, {Jacobs}, {Soenen},
  {Poedts}, {van der Holst}, \& {Zuccarello}}]{Zuccarello09}
{Zuccarello}, F.~P., {Jacobs}, C., {Soenen}, A., {et~al.} 2009, \aap, 507, 441

\end{thebibliography}

\end{document}